\documentstyle[a4,11pt]{article}

\input{psfig}
\def\Journal#1#2#3#4{#1 {\bf #2}, #3 (#4)}
\def\Book#1#2#3{{\it #1}, (#2, #3)}
\def\CMP{Commun. Math. Phys.}
\def\NPH{Nucl. Phys.}

\def\be{\begin{equation}}
\def\ee{\end{equation}}
\def\ba{\begin{array}}
\def\ea{\end{array}}
\def\bea{\begin{eqnarray}}
\def\eea{\end{eqnarray}}
\def\ben{\begin{eqnarray*}}
\def\een{\end{eqnarray*}}
\def\ds{\displaystyle}
\def\sm{\mbox{\small-}}
\def\sp{\mbox{\footnotesize+}}
\def\id{{\bf 1}}
\def\zero{{\bf 0}}

\def\nn{\nonumber}
\def\re{{\sf Re}}

\def\r{\mbox{\sf\ R}}
\def\z{\mbox{\sf Z}}
\def\c{\mbox{\sf C}}
\def\perm{\mbox{\footnotesize perm}}
\begin{document}

\thispagestyle{empty}
\setcounter{page}{0}
\vspace*{.6truein}

\centerline{\LARGE \bf Analytic Continuation of Operators ---}
\vskip .1truein
\centerline{{\LARGE \bf Operators acting complex s-times}}
\vskip .1truein
\centerline{{\bf Applications: from Number Theory and Group Theory}}
\centerline{{\bf to Quantum Field and String Theories}
\footnote{Preprint DAMTP-R-97/33 on-line at 
http://www.damtp.cam.ac.uk/user/scw21/papers/}}

\vskip .5truein
\centerline{S.C. Woon}
\vskip .2truein
\centerline{Department of Applied Mathematics and Theoretical Physics}
\centerline{University of Cambridge, Silver Street, Cambridge CB3 9EW,
  UK}
\vskip .1truein
\centerline{Email: S.C.Woon@damtp.cam.ac.uk}
\vskip .5truein

\centerline{\bf Abstract}

We are used to thinking of an operator acting once, twice, and so on.
However, an operator acting integer times can be consistently analytic
continued to an operator acting complex times. Applications:  (s,r)
diagrams and an extension of Fractional Calculus where commutativity of
fractional derivatives is preserved, generating integrals and non-standard
derivations of theorems in Number Theory, non-integer power series and
breaking of Leibniz and Chain rules, pseudo-groups and symmetry deforming
models in particle physics and cosmology, non-local effect in analytic
continued matrix representations, particle-physics-like scatterings of
zeros of analytic continued Bernoulli polynomials (physics/9705021),
analytic continuation of operators in QM, QFT and Strings.

\vfil \eject

\section{Introduction:\\ Analytic continuation of Operators}

\subsection{Questions in Context}

\begin{itemize}
\item What is $\ds \frac{d^{(1/2 + i)}}{dx^{(1/2 + i)}}\,?$ 
 Is it meaningful? What about $\ds \int (dx)^{(1/2 + i)}\;\,?$

\item Given a function $f(x)$, how do we evaluate $$\frac{d^{(1/2 +
 i)}}{dx^{(1/2 + i)}}\,f(x) \;\quad\mbox{and}\quad \int f(x)\,(dx)^{(1/2
 + i)}\;\;?$$ What do they mean?

\item Are these useful? Are there examples of applications?

\item If $A$ is a generic operator, how do we compute $A^{(1/2+i)}\,?$
  
\item For the creation and annihilation operators in Quantum Mechanics
  and Field Theories, how do we calculate and interpret
  $a^{(1/2+i)}\,|n\big>$ and the commutator
  $[\,a^{(1/2+i)},\,{(a^\dag)}^{(1/2+i)}\,]\;\,?$\\
  
\item What about similar generalisation to other operators, eg.
  Supersymmetric operators, Vertex operators, Virasoro algebra in
  String Theories, and Superconformal algebra in Superstring?

\item What are their surprising implications and consequences?

\end{itemize}

The aim of this paper is to address these issues and questions.

\subsection{The Idea}

We know that in Complex Analysis \cite{ahlfors}, functions can be
analytic continued from integer points $n$ on the real line to complex
plane $s$, eg. from fractorial $n!$ to Gamma function $\Gamma(s)$, and
from within the area of convergence to beyond via functional
equations, eg. the functional equation of the Riemann zeta function
\be\zeta(s) = 2^s\pi^{s-1}\sin\!\left(\frac{\pi s}{2}\right)
\Gamma(1-s)\;\zeta(1-s)\quad\;\forall\,s
\label{zeta_func_eqn})\ee

We too know that in Euclidean Quantum Gravity \cite{qg}, spacetime can be
analytic continued from
$$\mbox{Lorentzian spacetime }\to\mbox{ Complex spacetime }
\to\mbox{ Euclidean spacetime}$$
by rotating the signature of spacetime metric from
\be(-1,\sp 1,\sp 1,\sp1)\;\to\;(e^{i\,\theta},\sp 1,\sp 1,\sp
1)\;\to\;(\sp 1,\sp 1,\sp 1,\sp 1)\ee
Path integrals ill-defined in Lorentzian spacetime become well-behaved
in Euclidean spacetime in which they can be evaluated by methods of
infinite-descent and saddle-point approximation. This gives us a tool
to explore non-perturbative, non-linear and topological structures
like instantons and wormholes, the Thermodynamics of Hawking radiation
\cite{hawking_rad}, the end state of black hole evaporation \cite{evap},
and Conjectures on the boundary conditions of Complex spacetime
\cite{report}.

Now, we take analytic continuation a step further.

Can operators be analytic continued?

Operators and their representations permeate almost every branch of
Mathematics and field of Sciences. If the analytic continuation of
operators can be consistently defined and computed, then the idea may
have broad applications and universal implications.

We begin the first step with the two most familiar operators of all,
the differential operator $\ds \frac{d}{dx}$ and integral operator
$\ds \int dx\,$.

\section{Analytic continuation of $\ds\frac{d}{dx}$ and $\ds\int dx$}

\subsection{Differentiating and Integrating in non-integer $s$-dimensions}

Analytic continuation of differentiation and integration to
non-integer dimensions is straightforward.

Differential of an {\em integer} $n$-dimensional function
in $n$-dimensions is
\be\frac{\partial}{\partial x_1}\,\frac{\partial}{\partial x_2}\cdots
\frac{\partial}{\partial x_n}\,f(x_1,x_2,\cdots,x_n)\ee
and the corresponding integral is
\be\!\int f(x_1, x_2, \cdots, x_n)\,d^n x = \underbrace{ \int^{x_n}
  \cdots\int^{x_2} \int^{x_1}}_{n \mbox{\footnotesize-times}} f(x_1, x_2,
\cdots, x_n)\,dx_1 dx_2 \cdots dx_n\ee
If $f$ is spherically symmetric, $f = f(r)$, then\\
\be\frac{\partial}{\partial x_1}\,\frac{\partial}{\partial x_2}\cdots
\frac{\partial}{\partial x_n}\,=\,\frac{\partial}{\partial r}
\!\!\!\!{{}^{n-1}\atop{}^{\;\;n-1}}\,
\frac{\partial}{\partial\Omega}_{\!n-1}\!=\,
\frac{\Gamma(n/2)}{2\pi^{n/2}}\,
\frac{\partial}{\partial r}\!\!\!\!{{}^{n-1}\atop{}^{\;\;n-1}}\ee
can then be analytic continued to differential of a {\em non-integer}
$s$-dimensional function in $s$-dimensions as
\be\frac{\partial}{\partial r}\!\!\!\!{{}^{s-1}\atop{}^{\;\;s-1}}\,
\frac{\partial}{\partial\Omega}_{\!s-1}\!=\,
\frac{\Gamma(s/2)}{2\pi^{s/2}}\,
\frac{\partial}{\partial r}\!\!\!\!{{}^{s-1}\atop{}^{\;\;s-1}}\ee
and the corresponding integral 
\bea\int d^n x&=&\int^{\infty}_0 r^{n-1} dr \int^{2\pi}_0 d\theta_1
\int^\pi_0 \sin \theta_2 d\theta_2 \cdots \int^\pi_0 \sin^{n-2}
\theta_{n-1} d\theta_{n-1}\nn\\
 &=&\int^{\infty}_0 r^{n-1} dr \int d\Omega_{n-1}\,=\,
\frac{2\pi^{n/2}}{\Gamma(n/2)} \int^{\infty}_0 r^{n-1} dr\eea
as
\be\int d^s x\,=\,\frac{2\pi^{s/2}}{\Gamma(s/2)}\int^{\infty}_0
r^{s-1} dr\ee
where $s$ can be real or complex.

An important application example of such analytic continuation is
't Hooft and Veltman's Dimensional Regularization where it is used to
isolate singularities in divergent integrals in Quantum Field Theory
\cite{hooft}.

However, this is not the {\em only} possibility.

There is another possible analytic continuation.

\subsection{Differentiating and Integrating complex $s$-times
 in one-dimension}

Think of the differential and integral as {\em
  operators}. Differentiating or integrating $n$-times a
one-variable function $f(x)$ can be thought of as letting the operator
act $n$-times or $n$-fold on the function,
\be \left(\frac{d}{dx}\right)^{\!n} :\;f(x)\,\mapsto\,
\frac{d^n}{dx^n}\,f(x)\ee
$$\left(\int^x(d\hat{x})\right)^{\!n} :\;f(\hat{x})\,\mapsto\int^x\!
f(\hat{x})\,(d\hat{x})^n$$
\be = \underbrace{ \int^{x}\!\int^{x_n}\!\!\int^{x_{n-1}}\!\!\!\!\!\!\!\!\cdots
\!\int^{x_3}\!\!\int^{x_2}\!}_{n \mbox{\footnotesize-times}} f(x_1) \;
dx_1\,dx_2 \cdots dx_{n-1}\,dx_n\ee
Note that the limits of integration of this analytic continuation are 
different from those of integrating in $n$-dimensions.
\begin{itemize}
\item Integration in $n$-dimensions is a product of integrals

$\quad$ whereas

\item Integrating $n$-times gives a set of $n$ nested integrals with
  the limits taken at the end of the integration.
\end{itemize}
At this point, it is natural to generalise and combine both the
differential and integral operators into one fundamental operator
\be D^s_{\!x} = \left\{\ba{lcl}
\ds\frac{d^s}{dx^s}&,&\re(s)>0 \\
\ds\Big.{}\;1&,&\re(s)=0 \\
\ds\int (dx)^{-s}&,&\re(s)<0
\ea\right.\ee

The analytic continuation of the differential and integral operators
to the $D^s$ operator is known as Fractional Calculus
\cite{frac_calculus}, a subject of active research and of current
interest because of its widespread pure and applied applications in
Maths, Physics, Engineering and other Sciences \cite{frac_physics}.

In this paper, an extension of the conventional Fractional Calculus is
introduced.  Evaluating $\,D^s_{\!x} x^w\,$ is simple and obvious for
$\,\re(w)\ge 0\,$. On the other hand, for the case of
$\,\re(w)<0\,,\,$ it is not so straightforward but will turn out to be
simple when mapped to the $(s,w)$ diagrams to be introduced.

The surprise is that the consequences of analytic continuation of
these operators are not only highly non-trivial but useful.

In particular, a number of known and new results in Number Theory are
derived in non-standard ways using the idea of analytic continuation
of operators in Sections \ref{sec_num_th} and \ref{sec_op_series}.
These results demonstrate the usefulness and justify the purpose of
the idea.

In addition, when $D^s$ acts on a standard power series, the result is
a non-integer power series.  Analysis in Section
\ref{sec_nonint_pow_series} shows that there are interesting relations
between non-integer power series and the usual integer power series in
limiting forms. In fact, Fractional Calculus can be reinterpreted as
differential and integral operators acting non-integer times. $D^s$ is
observed to break Leibniz rule and Chain rule when $s$ is non-integer,
and thus we are unable to evaluate directly the action of $D^s$ on a
function of functions. However, by a trick of series expansion, we can
express $D^s$ as a nested sum of $d^n\!/dx^n$ or $\int (dx)^n$ which
we can evaluate directly with Leibniz rule and Chain rule.

Existing concepts in Group Theory are then extended in Section
\ref{sec_group} using these results. An extension of Dirac Algebra
from the Dirac equation is found in Section \ref{sec_algebra}. In
Finite Difference, the matrix representations of $d^n\!/dx^n$ and
$\int (dx)^n$ are sparse.  When the operator $D^s$ is casted in matrix
representation as in Section \ref{sec_fin_diff}, the matrix becomes
dense for non-integer $s$, and so the local finite difference becomes
non-local.  The cause of this non-local effect can be traced to the
appearance of non-integer power series.

Towards the end of the paper, problems were raised and challenges were
posed on analytic continuation of operators and algebras in Quantum
Mechanics, Supersymmetry, and Quantum Field and String Theories.

All in all, the analytic continuation of operators turns out to be
quite a general and powerful tool to explore Number Theory, Group
Theory, Algebra, Finite Difference and Matrix Representation.
Exploring with the idea has motivated the introduction of a few other
new ideas and concepts into these fields, each is of very different
nature from the other. The intriguing results suggest that the idea
of analytic continuation of operators may well find interesting
widespread applications in various other fields.

\section{Defining the action of operator $D^s$}

The operator $D^n_{\!x}$ for integer $n$ is well-defined since it
corresponds to $\ds \frac{d^n}{dx^n}$ for $n \ge 1$ and $\ds \int
(dx)^{-n}$ for $n \le -1$. 
$$\left\{\ba{l}
\ds\frac{d^n}{dx^n}\,x^m = \left\{
\ba{lcl}{}\,0&\quad,&0<m\le n\\
\ds\frac{m!}{(m\!-\!n)!}\,x^{m-n}&\quad,&m>n>0\ea\right.\\
\\
\ds\int x^m\,(dx)^{-n} \,=\, \ds\frac{m!}{(m\!-\!n)!}\,x^{m-n}\;\;,\;\;
m>0\;\;,\;n<0
\ea\right\},\;m,n\in\z$$
If we tabulate $D^n_{\!x} x^m$ for
integers $n,m\,$, we observe a pattern emerges.
\begin{table}[hbt]
\caption{Tabulated results of $D^n_{\!x} x^m$}
\bigskip
$\!\!\!\!\begin{array}{|r||r|r|r|r|r|r|r|}
\hline
  \multicolumn{8}{|c|}{D^n_{\!x} x^m} \\
  \hline
  \mbox{\small\it m}\!\backslash\!\mbox{\small\it n} & -3 & -2 & -1 & 0
  & 1 & 2 & 3 \\
  \hline \hline 2 & \mbox{\small 2!/5!}\;x^5 & \mbox{\small
    2!/4!}\;x^4 & \mbox{\small 2!/3!}\;x^3 & x^2 & \mbox{\small 2!}\;x
  & \mbox{\small 2!} & \mbox{\small 0}  \\
  \cline{7-7} 1 & \mbox{\small 1/4!}\;x^4 & \mbox{\small 1/3!}\;x^3 &
  \mbox{\small 1/2!}\;x^2 & x & \mbox{\small 1} & \mbox{\small 0} &
  \mbox{\small 0} \\
  \cline{6-6} 0 & \mbox{\small 1/3!}\;x^3 & \mbox{\small 1/2!}\;x^2 &
  x & \mbox{\small 1} & \mbox{\small 0} & \mbox{\small 0} & \mbox{\small 0} \\
  \cline{2-4} \cline{6-8} -1 & \int\log x\,(\!\mbox{\small\it dx})^2 &
  \int\log x\,(\!\mbox{\small\it dx}) & \log x & x^{-1} &
  \mbox{-}\,x^{-2} & \mbox{\small 2!}\;x^{-3} &
  \mbox{-\small 3!}\;x^{-4} \\
  \cline{4-4} -2 & \mbox{-}\!\int\log x\,(\!\mbox{\small\it dx})
  & \mbox{-}\log x & \mbox{-}\,x^{-1} & x^{-2} &
  \mbox{-\small 2!}\;x^{-3} & \mbox{\small 3!}\;x^{-4} &
  \mbox{-\small 4!}\;x^{-5} \\
  \cline{3-3} -3 & \mbox{\small 1/2!}\; \log x & \mbox{\small 1/2!}\;
  x^{-1} & \mbox{-\small 1/2!}\;x^{-2} & x^{-3} & \mbox{-\small
    3!/2!}\;x^{-4} & \mbox{\small 4!/2!}\;x^{-5} &
  \mbox{-\small 5!/2!}\;x^{-6}\\
  \hline
\end{array}$
\label{D_table}
\end{table}

\subsection{Fractional Calculus and its Extension}

There are several possible ways for analytic continuing $d/dx$ and
$\int dx\;$:

\begin{enumerate}
\item {\em Riemann-Liouville Fractional Calculus}

In the conventional Riemann-Liouville Fractional Calculus
\cite{frac_calculus}, we start with the Riemann-Liouville Integral
\be \underbrace{\int^{x}_{\!a}\!\int^{x_n}_{\!a}\!\!
\int^{x_{n-1}}_{\!a}\!\!\!\!\!\!\!\!\cdots\!\int^{x_3}_{\!a}
\!\!\int^{x_2}_{\!a}\!}_{n \mbox{\footnotesize-times}} f(x_1) \;
dx_1\,dx_2 \cdots dx_{n-1}\,dx_n = \frac{1}{\Gamma{(n)}}
\int_a^{x}\!\!\frac{f(t)}{(x - t)^{1-n}}\,dt\label{R_L_int}\ee
as the fundamental defining expression.

Fractional integral is analytic continued from this Riemann-Liouville
integral (\ref{R_L_int}) as
\bea D^\sigma_{\!x-a} f(x)&=&\frac{d^\sigma}{d(x-a)^\sigma}
f(x)\;\,=\;\,\int_a^x f(x) (dx)^{-\sigma}\nn\\
&=&\frac{1}{\Gamma(-\sigma)}\int_a^x\!\!\frac{f(t)}{(x-t)^{1+\sigma}}
\;dt\quad\mbox{ for }\;\;\sigma<0,\;\,\sigma,a\in\r\label{frac_int}\eea

Fractional derivative is in turn derived from the fractional integral
(\ref{frac_int}) by ordinary differentiation as $\;$ (choose $m>\sigma$)
$$D^\sigma_{\!x} f(x)\,=\,D^m_{\!x} \left( D_{\!x}^{-(m-\sigma)} f(x)
\right)\quad\mbox{ for }\;\;\sigma>0,\;\,m\in\z^+$$
and it has the property
$$D^\sigma_{\!x} f(\beta x)\,=\,\beta^\sigma\,D^\sigma_{\!(\beta x)}
f(\beta x)$$

However, for $\,D^\sigma_{\!x} x^r$, the Riemann-Liouville Integral
definition is well-defined only for the half plane
$\,\sigma\in\r\,,\,r>-1$.

\item {\em Fractional Calculus by Cauchy Integral}

Cauchy Integral for an analytic function $f(z)$ in the complex plane is
\be f^{(n)}(z_0)\,=\,\frac{\Gamma(1+n)}{2\pi i}\int_C
\frac{f(z)}{(z-z_0)^{1+n}}\,dz \ee
Generalization of $\,n\,$ to non-integer values is however not trivial
as the term $(z-z_0)^{1+\alpha}$ may become multi-valued and the result
will depend on the choice of branch cut and integration path.

\item {\em Fractional Calculus by Fourier Transform or other Integral
    Transform}

In Fourier Transform,
\be \tilde{f}(x)\,=\,\int_{-\infty}^{+\infty}f(x)\,e^{ikx}\,dx\quad,\quad
f(x)=\frac{1}{2\pi}\,\int_{-\infty}^{+\infty}\tilde{f}(x)\,e^{-ikx}\,dk\ee
\bea D^\sigma_{\!x} f(x)&=&\int_{-\infty}^{+\infty}\tilde{f}(k)\,
D^\sigma_{\!x}\!\left(e^{-ikx}\right)dk\;\;,\quad\sigma\in\r\nn\\
&=&\int_{-\infty}^{+\infty}(-ik)^\sigma\tilde{f}(k)\;e^{-ikx}\,dk\nn\\
\Rightarrow\;\;D^\sigma_{\!x}&\equiv&\int_{-\infty}^{+\infty}(-ik)^\sigma
e^{-ikx}\,dk\eea
This was shown by Z\'avada \cite{frac_fourier} to be equivalent to the
Riemann-Liouville Fractional Calculus and the Fractional Calculus by
Cauchy Integral.

Now, to computed numerically the fractional derivative or fractional
integral of a function $f(x)$ or multi-variable function
$f(x^{(1)},x^{(2)},\cdots,x^{(n)})$, we choose to use the Discrete
Fast Fourier Transform (DFFT) instead.  Take the function values in
the interval or region of interest, identify the boundaries so as to
make it periodic, feed this into the DFFT algorithm to Fourier
transform the function into $k$-space, multiply the component
corresponding to $k$ with $(-ik)^\sigma$, and feed the result into the
Inverse DFFT to Inverse Fourier transform back it to $x$-space.

\item {\em Extended Fractional Calculus}

Now, we introduce the Extended Fractional Calculus in which
the limit of the ratio of Gamma functions
\be\lim_{\epsilon\to 0}\,\frac{\Gamma(1\!+\!r\!+\!\epsilon)}
{\Gamma(1\!+\!r\!+\!\epsilon\!-\!\sigma)}\ee
is taken as the fundamental defining expression instead.

Unlike the Riemann-Liouville Integral, the limit is well-defined for
the entire plane $\,{\sigma,r\in\r}\,$ except along the line intervals
$\,{r\in\z^-,\,\sigma\in\r\big\backslash\z}$.  The analytic
continuation to these line intervals will be derived in Section
\ref{sec_lower_log_region}, equation (\ref{D_horz_intv}).

However, in the wedge-shaped region $\,r<\sigma\,,\,r\ge 0$, there are
actually two possible choices, of which one ({\em Type I$\,$})
corresponds to the Riemann-Liouville Fractional Calculus in that
region where fractional derivatives do not generally commute, and the
other ({\em Type II$\,$}) an extension of it where commutativity of
the fractional derivatives is preserved. This will be described in
Section \ref{sec_zero_region}.

\item {\em Fractional Calculus by Nested Series Expansion}

Here, we introduce the method of analytic continuation of operators by
 nested series expansion with the following observation:

$D^s$ can be formally expanded into a nested series as
\bea 
\!\!\!\!D^s &=& \left( \Big. w\id - \Big[ w\id - D \Big]
   \right)^{\!s} \;=\;w^s\!\left(\id-\Big[\id-\frac{1}{w}\,D\Big]
\right)^{\!\!s}\nn\\ 
&=& w^s \left(\id + \sum_{n=1}^\infty \frac{(-1)^n}{n!} \left[
    \prod_{k=0}^{n-1} (s \!-\! k) \right] \!\left[ \Big.\id -
    \frac{1}{w}\,D \right]^{\!n} \right) \nn\\ 
&=& w^s\left( \id + \sum_{n=1}^\infty \frac{(-1)^n}{n!} \left[
    \prod_{k=0}^{n-1} (s \!-\! k) \right] \!\!\left[\,\id + \sum_{m=1}^n 
     \!\left(\frac{-1}{w}\right)^{\!\!m} \!\! {n \choose m} D^m \right]
      \right)
\label{D_nested_series}
\eea
where $\;s,w\in\c$, and $\;\id\;$ is the identity operator.

In the nested series on {r.h.s}, all the operators $D$'s are raised to
integer powers $m$, and $D^m$ corresponds to ordinary $m$-fold
differentiation if $\,{m<0}\,$ or integration if $\,{m>0}\,$.  The
region of convergence in $\,s\,$ and the rate of convergence of the
series will be dependent on parameter $w,\;$ and the function on which
$D$ acts.

In this way, Fractional Calculus with its analytic continued operator
$D^s$ can be reinterpreted as operator $D$ acting non-integer or
complex $s$-times on a function. Examples of applications can be found
in Sections \ref{sec_breaking_rules} and \ref{sec_problems}, and in
\cite{woon_bernoulli_con}.

\end{enumerate}

\subsection{Analytic continuation from integer $n$ to real $\sigma$}

\begin{figure}[hbt]
\begin{center}
\begin{tabular}{c}
\psfig{figure=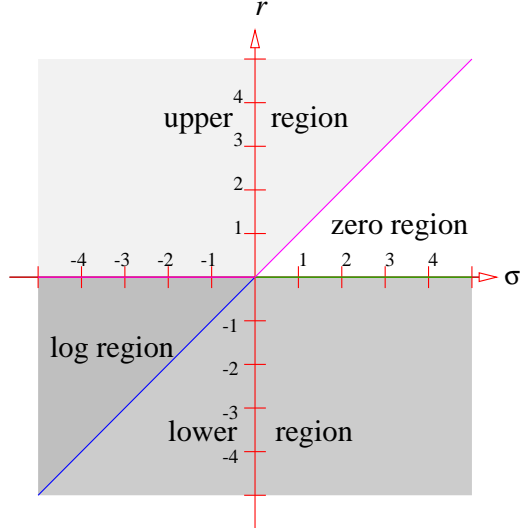,height=200pt}
\end{tabular}
\caption{$(\sigma,r)$ diagram of $D^\sigma_x x^r$.\label{sigma_r_diagram}}
\end{center}
\end{figure}

To analytic continue to $D^\sigma_{\!x} x^r$ for real $\sigma,r$, we
introduce the $(\sigma,r)$ diagram in which the coefficient of
$D^\sigma_{\!x} x^r$ is mapped to the point at coordinate $(\sigma,r)$
of the diagram. The $(\sigma,r)$ diagram of $D^\sigma_{\!x} x^r$ can
then be characterized into 5 regions as in Figure \ref{sigma_r_diagram}:\\
\be\left.\ba{c}\mbox{upper}\\\mbox{lower}\\\mbox{log}\\\mbox{zero}\ea
\right\}\mbox{ region bounded by }\left\{
\ba{l}r\ge\sigma\;,\;\;r\ge0\\
r<\sigma\;,\;\;r<0\\
r\ge\sigma\;,\;\;r<0\\
r<\sigma\;,\;\;r\ge0
\ea\right.\ee
$$\mbox{A point lying on the } \left\{ \ba{cll}\mbox{right} 
 &(\sigma>0)& \mbox{ is a differentiation}\\
\mbox{left } &(\sigma<0)& \mbox{ is an integration} \ea\right.$$

\subsubsection{Upper region}

\be\mbox{The ratio }\quad
\frac{\Gamma(1\!+\!r)}{\Gamma(1\!+\!r\!-\!\sigma)}\qquad\qquad\qquad\qquad
\qquad\qquad\qquad\qquad\qquad\qquad\label{gamma_ratio}\ee is finite
everywhere in $\;r\ge\sigma\,,\;r\ge 0$.

So,\\
\indent$\ds\quad\quad D^\sigma_{\!x} x^r = 
 \frac{\Gamma(1\!+\!r)}{\Gamma(1\!+\!r\!-\!\sigma)}\,
x^{r-\sigma}\;\;$ in the upper region.

\subsubsection{Lower and Log regions\label{sec_lower_log_region}}

Define

$\Omega_{r<0}$ as the union of lower and log regions such that 
$\,r<0\,,\sigma\in\r\,,$

$\Omega_{hor\!z}$ as the set of the horizontal lines $\,r =
-n_+\,,\;n_+\!\in\z^+$,

$\Omega_{grid}$ as the set of integer grid points in $\Omega_{r<0}$
such that $\,\sigma\in\z\,,\;r\in\z^-$,

$\Omega_{diag}$ as the set of right-sloping diagonals in the lower
region\\
\indent$\qquad\qquad r=\sigma-n_+\!<0\,,\;n_+\!\in\z^+\,,\;\sigma,r\in\r\,$.
\begin{figure}[hbt]
\begin{center}
\begin{tabular}{c}
\psfig{figure=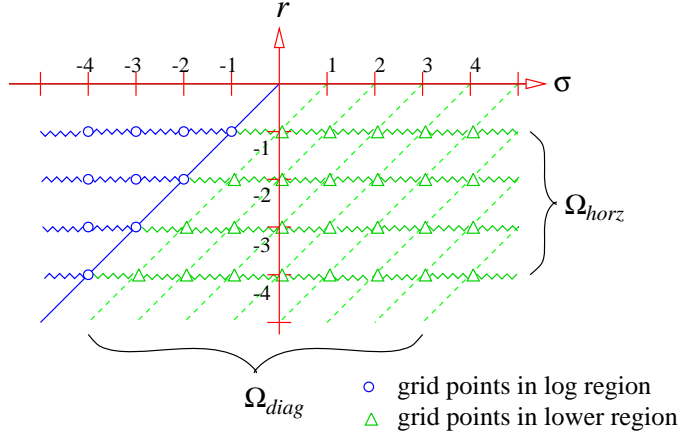}
\end{tabular}
\caption{Lower and Log regions.}
\end{center}
\end{figure}

The ratio (\ref{gamma_ratio}) is finite everywhere in $\Omega_{r<0}$
{\em except} in $\Omega_{hor\!z}\!\Big\backslash\Omega_{grid}\,$
(along the horizontal lines ``mod'' the grid points). It is zero in
$\,\Omega_{diag}\!\Big\backslash\Omega_{grid}\,$ (along the
diagonals in the lower region ``mod'' the grid points).

However, the limit 
\be\ds\qquad\quad\lim_{\epsilon\to 0} \,
\frac{\Gamma(1\!+\!r\!+\!\epsilon)}
{\Gamma(1\!+\!r\!+\!\epsilon\!-\!\sigma)}
\qquad\qquad\qquad\qquad\qquad\qquad\qquad\qquad\qquad\ee
evaluated at the grid points in $\Omega_{grid}$ is convergent.

Thus,
\be D^\sigma_{\!x} x^r = \lim_{\epsilon\to 0} \,
\frac{\Gamma(1\!+\!r\!+\!\epsilon)}
{\Gamma(1\!+\!r\!+\!\epsilon\!-\!\sigma)}\,x^{r-\sigma}\;\;
\mbox{ in }\;\Omega_{r<0}\!\Big\backslash\!\!\left(\Omega_{hor\!z}
\!\Big\backslash\Omega_{grid}\right)\quad
\label{D_lower_log}\ee

From Table \ref{D_table}, the natural analytic continuation of
$D^\sigma_{\!x} x^r$ in the part of
$\Omega_{hor\!z}\!\Big\backslash\Omega_{grid}$ lying in the log region is
\be \lim_{\epsilon\to 0}\,
\frac{\Gamma(1\!+\!r\!+\!\epsilon)}{\Gamma(\epsilon)} \int\log
x\,(dx)^{(r-\sigma)} \equiv \;\lim_{\epsilon\to 0}\,
\frac{\Gamma(1\!+\!r\!+\!\epsilon)}{\Gamma(\epsilon)}\,D^{(\sigma-r)}_{\!x}
\log x \label{D_omega_expression}\ee

It can be shown using (\ref{D_lower_log}) and the following expression
\be\log x = \log x + \log 1 = \lim_{\epsilon \to 0} \int_1^x
\hat{x}^{-1+\epsilon} \, d\hat{x} = \lim_{\epsilon \to 0}
\frac{1}{\epsilon}\,( x^\epsilon - 1 ) \label{log_x}\ee
that
\be\int\log x\,(dx)^\rho\,\equiv\,D^{-\rho}_{\!x}\log x\,=\,
\frac{x^\rho}{\Gamma(1\!+\!\rho)} \left(\Big.\log x -
  (\big.\psi(1\!+\!\rho)+\gamma)\right) \label{log_psi}\ee
where the digamma function \cite{math_defn}
$$\psi(z) = \frac{d}{dz} \log\left(\Gamma(z)\right) =
\frac{\Gamma'(z)}{\Gamma(z)}$$
the Euler constant $\;\gamma = -\,\Gamma'(1) = 0.577215\cdots$, and
the prime ' denotes differentiation.  This will be proved in
Section \ref{sec_num_th}, equation (\ref{harmonic}).

With (\ref{log_psi}), expression (\ref{D_omega_expression}) can then
be analytic continued from the part of
$\Omega_{hor\!z}\!\Big\backslash\Omega_{grid}$ lying in the log region
into that lying in the lower region.

Thus, in $\Omega_{hor\!z}\!\Big\backslash\Omega_{grid}$, we have
$\,r\in\z^-,\;(\sigma\!-\!r)\not\in\z\,,\,$ and the analytic
continuation as
\bea D^{\sigma}_{\!x} x^r&=&
\lim_{\epsilon\to 0}\,
\frac{\Gamma(1\!+\!r\!+\!\epsilon)}{\Gamma(\epsilon)}\,D^{(\sigma-r)}_{\!x}
\log x\nn\\
&=&\lim_{\epsilon\to 0}\,
\frac{\Gamma(1\!+\!r\!+\!\epsilon)}{\Gamma(\epsilon)}\,
\frac{x^{(r-\sigma)}}{\Gamma(1\!+\!r\!-\!\sigma)} \left(\Big.\log x -
  (\big.\psi(1\!+\!r\!-\!\sigma)+\gamma)\right)\label{D_horz_intv}\eea

\subsubsection{Zero region\label{sec_zero_region}}

As pointed out above, there are two possible choices in this region
from the two views or schools of thoughts:

\begin{itemize}
\item {\em Type I $\,$ Fractional Calculus}

Postulate:\\
Fractional derivative is abstract. Fractional derivative of a constant
can be non-zero. So take the Riemann-Liouville Integral as fundamental
and derive the fractional derivative in this zero region from it.

Then,
\be D^\sigma_{\!x} x^r\,=\,\frac{\Gamma(1\!+\!r)}
{\Gamma(1\!+\!r\!-\!\sigma)}\quad
\mbox{in the zero region.}\qquad\qquad\qquad\qquad\ee

\item {\em Type II $\,$ Fractional Calculus}

Postulate:\\
Ordinary derivative of a constant is zero and ordinary derivatives
commute. Fractional derivative should inherit these property from
ordinary derivative as well --- fractional derivative of a constant is
zero and fractional derivates commute.
\bea D^1_{\!x}\,c&\!\!=&\!\!0\;\;,\quad c \mbox{ is an arbitrary
  constant}\nn\\
D^\sigma_{\!x}\,c&\!\!=&\!\!D^{(\sigma-1)}_{\!x}\!
\left(D^1_{\!x}\,c\right)\,=\,D^{(\sigma-1)}_{\!x}\;0\;=\;0\;\;,\quad
\mbox{for }\;\sigma>1\qquad\quad\eea
As

\noindent$\qquad\;\;D^r_{\!x}x^r\,=\,\Gamma(1+r)\quad\mbox{ for }\;r\ge0$

by continuity, 
$$D^\sigma_{\!x} x^r\,=\,0\quad\mbox{ for }\;\sigma>r\,,\,r\ge0
\qquad\qquad\qquad\qquad\qquad\qquad\qquad$$
\noindent$\;\;\Rightarrow\;\; D^\sigma_{\!x} x^r\,=\,0\;\;$ in the
zero region and commutativity is preserved.
\end{itemize}

\noindent Hence, in both Types, there is a mutual trade-off.

\noindent In {\em Type I}$\,$, we chose analyticity and lose
commutativity in the zero region.

\noindent In {\em Type II}$\,$, we chose to preserve and carry over
commutativity and lose analyticity at the edges of the zero region.

\subsubsection{Entire real $(\sigma,r)$ plane}

The combined analytic continuation is then

\noindent$\bullet\;\;${\em Type I}
\be D^\sigma_{\!x} x^r = \left\{
\ba{ll}
\multicolumn{2}{l}{
\ds\Bigg.\lim_{\epsilon\to 0}\,
\frac{\Gamma(1\!+\!r\!+\!\epsilon)}{\Gamma(\epsilon)}\,
\frac{x^{(r-\sigma)}}{\Gamma(1\!+\!r\!-\!\sigma)} \left(\Big.\log x -
  (\big.\psi(1\!+\!r\!-\!\sigma)+\gamma)\right)}\\
&\qquad\qquad\qquad\qquad
\mbox{ in }\;\Omega_{hor\!z}\!\Big\backslash\Omega_{grid}\\
\\
\ds\Bigg.\lim_{\epsilon\to 0}\,
 \frac{\Gamma(1\!+\!r\!+\!\epsilon)}
{\Gamma(1\!+\!r\!+\!\epsilon\!-\!\sigma)}\,
x^{r-\sigma}&\qquad\qquad\qquad\qquad\mbox{ elsewhere}
\ea\right.\label{D_combined_I}\ee

\noindent$\bullet\;\;${\em Type II} $\;$ (fractional derivatives commute)
\be D^\sigma_{\!x} x^r = \left\{
\ba{ll}
\ds\Big.\,0&\qquad\qquad\qquad\qquad\mbox{ in the zero region}\\
\\
\multicolumn{2}{l}{
\ds\Bigg.\lim_{\epsilon\to 0}\,
\frac{\Gamma(1\!+\!r\!+\!\epsilon)}{\Gamma(\epsilon)}\,
\frac{x^{(r-\sigma)}}{\Gamma(1\!+\!r\!-\!\sigma)} \left(\Big.\log x -
  (\big.\psi(1\!+\!r\!-\!\sigma)+\gamma)\right)}\\
&\qquad\qquad\qquad\qquad
\mbox{ in }\;\Omega_{hor\!z}\!\Big\backslash\Omega_{grid}\\
\\
\ds\Bigg.\lim_{\epsilon\to 0}\,
 \frac{\Gamma(1\!+\!r\!+\!\epsilon)}
{\Gamma(1\!+\!r\!+\!\epsilon\!-\!\sigma)}\,
x^{r-\sigma}&\qquad\qquad\qquad\qquad\mbox{ elsewhere}
\ea\right.\label{D_combined_II}\ee

\subsection{Analytic continuation from real $\sigma$ to complex $s$}

To analytic continue $D^\sigma$ to $D^s$ where complex
${s=\sigma\!+\!it}$, we just generalise $\,\sigma\,$ to $\,s\,$ in the
combined expressions (\ref{D_combined_I}) and (\ref{D_combined_I}) for
for $D^s_{\!x} x^w$ with complex $s,w$.

For {\em Type II}$\;$ Fractional Calculus, commutativity is
preserved.  The differential operator $d/dx$ commutes with itself and
with its inverse, the integral operator $\int dx$, and so do $D^s$.

\begin{figure}[hbt]
\begin{center}
\begin{tabular}{c}
\psfig{figure=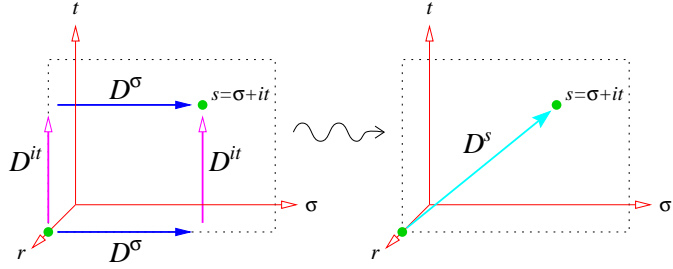,height=100pt}
\end{tabular}
\caption{Commutative arrows in $(s,r)$ diagram of $D^s_{\!x} x^r$ in
  {\em Type II}$\,$.
\label{D_commutative_diagram}}
\end{center}
\end{figure}

So in {\em Type II}$\,$, the commutative operator $D^s$ splits into two,
each acting on functions independently.  \be D^s = D^{(\sigma+it)} =
D^\sigma D^{it} = D^{it} D^\sigma\ee as illustrated in the commutative
diagram of Figure \ref{D_commutative_diagram}, with all the limits, if
any, taken {\em only at the end} after the actions of all the
operators have been performed.

Consider $D^s_{\!x}x^w$ with complex $s=\sigma\!+\!it\,,\;w=u\!+\!iv$.

In the zero region of $(\sigma,u)$ plane,

$\qquad D^\sigma_{\!x}x^u \;= 0 \quad\Rightarrow\quad D^{it}_{\!x}
D^\sigma_{\!x}x^u\,=D^s_{\!x}x^u\,=0$

Similarly, in the zero region of $(it,iv)$ plane,

$\qquad D^{it}_{\!x}x^{iv} = 0 \quad\Rightarrow\quad D^\sigma_{\!x}
D^{it}_{\!x}x^{iv} = D^s_{\!x}x^{iv} = 0$

\begin{figure}[hbt]
\begin{center}
\begin{tabular}{c}
\psfig{figure=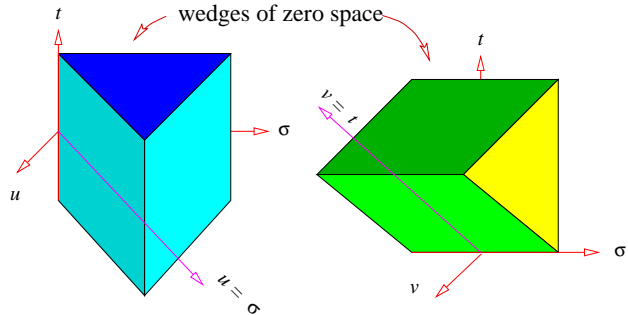,height=120pt}
\end{tabular}
\caption{The zero space wedges in $(s,w)$ diagrams where
  $D^s_{\!x} x^w$ in {\em Type II}$\,$. \label{rszero}}
\end{center}
\end{figure}

We can think of the above as $D^{it}$ expanding the triangular zero
region of the $(\sigma,u)$ plane into a wedge-shaped volume of
infinite length along the $t$-direction, and $D^{\sigma}$ similarly
expanding that in $(it,iv)$ plane into another along the
$\sigma$-direction as shown in Figure \ref{rszero}.

This defines the zero space in which $D^s_{\!x}x^w = 0\,$.

For the line intervals in
$\Omega_{hor\!z}\!\Big\backslash\Omega_{grid}\,,\;D^{it}$ extends
these intervals in the $(\sigma,u)$ plane along the $t$-direction to
planar sections, and similarly for $D^{\sigma}$.

By now, we have completed the task of defining the action and obtained
a complete picture of the operator $D^s$. In essence, what we have
done is analytic continuing an operator to act on a function complex
$s$-times. Let us carry on with applications of the analytic continued
$D^s$ to find out more about its properties and consequences.

\section{Application: Number Theory $-$\\
{\normalsize\bf Generating Integrals and\\
Analytic continuation of Finite Series}\label{sec_num_th}}
\begin{quote}{\em ``Nature laughs at the difficulties of
    integration.''} --- Laplace\end{quote}

\subsection{Generating Integral of Finite Harmonic Series}

It is well known \cite{math_defn} that finite harmonic
series
\bea h(n)&=&\sum_{k=1}^{n}\frac{1}{k} = 1 + \frac{1}{2} + \cdots +
\frac{1}{n}\nn\\
&=&\psi(1+n)+\gamma\\
&=&\log\,n + \gamma + O(1/n)\nn \eea

Now, we observe that finite harmonic series $h(n)$ appears as the
coefficient of $x^n$ term when we repeatedly integrate $\log x$.
\ben\int_0^x \log\hat{x}\,(d\hat{x}) &\!\!\!=&\!\!\! x (\, \log x - 1 \,)\\
\int_0^x \log\hat{x}\,(d\hat{x})^2 &\!\!\!=&\!\!\! \frac{x^2}{2} (\,
\log x -
\frac{3}{2} \,)\\
& \vdots &\\
\int_0^x \log\hat{x}\,(d\hat{x})^n &\!\!\!=&\!\!\! \frac{x^n}{n!} (\,
\log x - h(n) \,)\een
So by analogy to the concept of generating functions, we take\\
$$\int_0^x \log x\,(dx)^n \mbox{ as the {\em generating
    integral} of finite harmonic series } h(n)$$\\ 
and so the natural analytic continuation of the generating integral
takes the form of
\be\int_0^x \log\hat{x}\,(d\hat{x})^\rho =
\frac{x^\rho}{\Gamma(1\!+\!\rho)} \left( \log x - h(\rho) \right)
\label{intlog}\ee
From (\ref{log_x}),
$$\log\,x =\,\lim_{\epsilon\to0}\frac{1}{\epsilon}\,(x^\epsilon - 1 )$$
The analytic continuation of the integral on {l.h.s.} of
(\ref{intlog}) can then be evaluated as
\bea\int_0^x \log\hat{x}\,(d\hat{x})^\rho &\!\!\!=&\!\!\! \lim_{\epsilon
  \to 0} \frac{1}{\epsilon} \int_0^x (\hat{x}^\epsilon - 1)
  (d\hat{x})^\rho \nn\\ 
&\!\!\!=&\!\!\! \lim_{\epsilon \to 0} \frac{1}{\epsilon}
  \left[ \int_0^x \hat{x}^\epsilon (d\hat{x})^\rho - \int_0^x 1\,
  (d\hat{x})^\rho \right] \nn\\ 
&\!\!\!=&\!\!\! \lim_{\epsilon \to 0} \frac{1}{\epsilon}\,
  \bigg[ D^{-\rho}_{\!\hat{x}} \hat{x}^\epsilon \Big|_0^x \;\,-\;\,
  D^{-\rho}_{\!\hat{x}} \,1\, \Big|_0^x \bigg] \nn\\ 
&\!\!\!=&\!\!\! \lim_{\epsilon \to 0} \frac{x^\rho}{\epsilon} \left[ 
\frac{\Gamma(1\!+\!\epsilon)\,x^\epsilon}{\Gamma(1\!+\!\epsilon\!+\!\rho)} -
  \frac{1}{\Gamma(1\!+\!\rho)} \right]\eea

Equating this to the expression on {r.h.s.} of (\ref{intlog}) gives
the analytic continuation of finite harmonic series \bea h(\rho)
&\!\!\!=&\!\!\! \ds \log x - \lim_{\epsilon \to 0} \frac{1}{\epsilon}
\left[ \frac{\Gamma(1\!+\!\epsilon)\,
    \Gamma(1\!+\!\rho)}{\Gamma(1\!+\!\epsilon\!+\!\rho)}\,
x^\epsilon - 1\right]\nn\\
&\!\!\!=&\!\!\! \log x - \lim_{\epsilon \to 0} \frac{1}{\epsilon}
\left[ \ba{l} \ds \Bigg.
\frac{\Gamma(1\!+\!\epsilon)\,\Gamma(1\!+\!\rho)}
{\Gamma(1\!+\!\epsilon\!+\!\rho)}\,x^\epsilon -
\frac{\Gamma(1\!+\!\epsilon)\,\Gamma(1\!+\!\rho)}
{\Gamma(1\!+\!\epsilon\!+\!\rho)}\nn\\ \ds \Bigg. +
\frac{\Gamma(1\!+\!\epsilon)\,\Gamma(1\!+\!\rho)}
{\Gamma(1\!+\!\epsilon\!+\!\rho)}\;-\;1 \ea\right]\nn\\
&\!\!\!=&\!\!\! \ds \log x - \lim_{\epsilon \to 0}\,
\frac{1}{\epsilon}\,(x^\epsilon - 1) \left[ \frac{\Gamma(1\!+\!\epsilon)\,
    \Gamma(1\!+\!\rho)}{\Gamma(1\!+\!\epsilon\!+\!\rho)} \right]\nn\\
& & \ds+\;\lim_{\epsilon\to 0} \frac{1}{\epsilon} \left[ 1 -
  \frac{\Gamma(1\!+\!\epsilon)\,
    \Gamma(1\!+\!\rho)}{\Gamma(1\!+\!\epsilon\!+\!\rho)} \right]\nn\\
&\!\!\!=&\!\!\! \ds \left[\log x - \lim_{\epsilon \to 0}\,
\frac{1}{\epsilon}\,(x^\epsilon - 1)\right]
+\lim_{\epsilon\to 0}\frac{1}{\epsilon} \left[ 1 -
  \frac{\Gamma(1\!+\!\epsilon)\,
    \Gamma(1\!+\!\rho)}{\Gamma(1\!+\!\epsilon\!+\!\rho)} \right]\nn\\
&\!\!\!=&\!\!\! \ds \lim_{\epsilon \to 0} \frac{1}{\epsilon} \left[ 1
  - \frac{\Gamma(1\!+\!\epsilon)\,
    \Gamma(1\!+\!\rho)}{\Gamma(1\!+\!\epsilon\!+\!\rho)} \right]\nn\\
&\!\!\!=&\!\!\! \ds \lim_{\epsilon \to 0} \; - \, \Gamma(1\!+\!\rho)
\left[ \frac{\Gamma'(1\!+\!\epsilon\!+\!\rho)\,\Gamma(1\!+\!\epsilon) -
    \Gamma'(1\!+\!\epsilon)\,
    \Gamma(1\!+\!\epsilon\!+\!\rho)}{\Gamma(1\!+\!\epsilon\!+\!\rho)^2}
\right]\nn\\
\Bigg.&\!\!\!=&\!\!\!\ds\frac{\Gamma'(1\!+\!\rho)}{\Gamma(1\!+\!\rho)} -
\Gamma'(1)\nn\\
\Big.&\!\!\!=&\!\!\!\ds\psi(1\!+\!\rho) + \gamma\label{harmonic}\eea
where the limit $\epsilon \to 0$ has been taken with L'Hospital rule.

(\ref{harmonic}) is a result that has been proved in conventional ways in
Number Theory. It is somewhat surprising and miraculous that it is
possible too to rederive it with the method of the analytic continued
operator $D^s$ as above.
\begin{figure}[hbt]
\begin{center}
\begin{tabular}{c}
\raisebox{50pt}{$\!\!\!\!\!\!h(\rho)$}
\psfig{figure=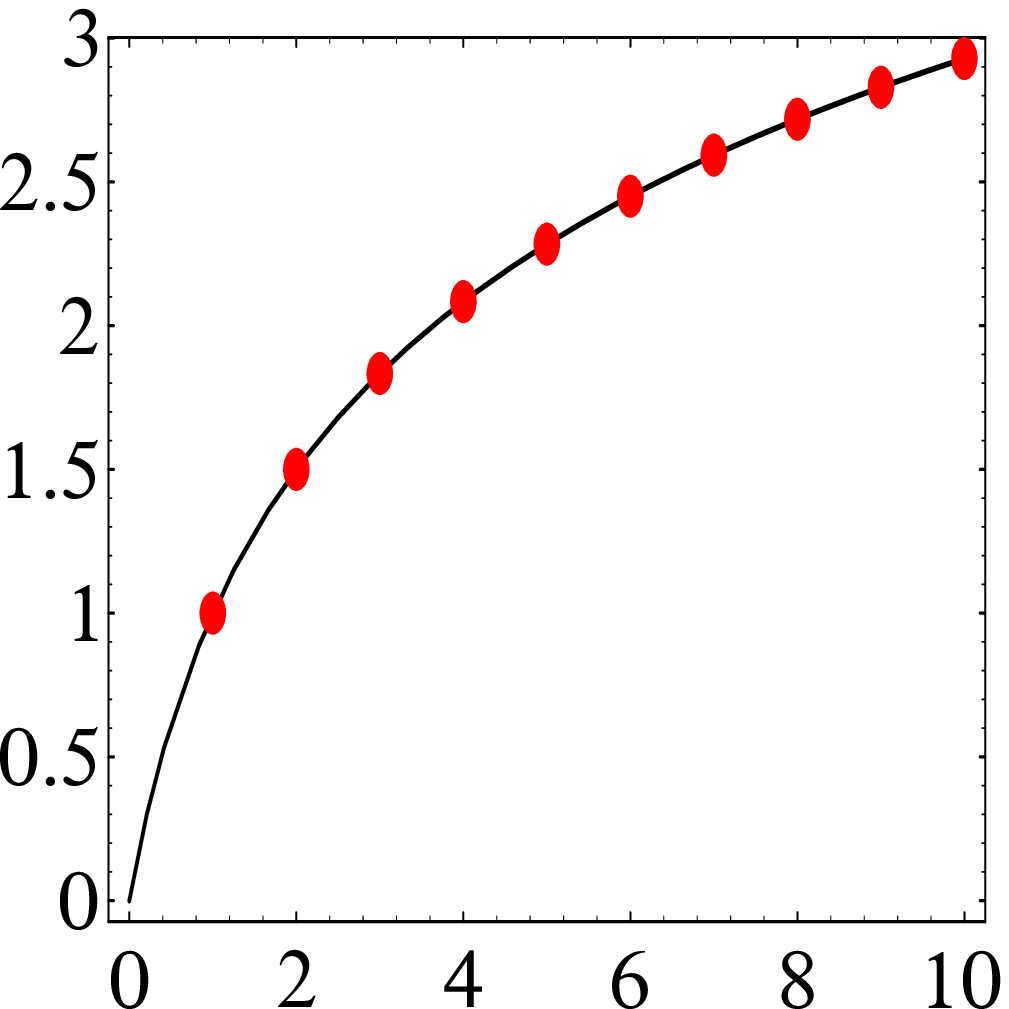,height=100pt}\\
$\qquad\rho$
\end{tabular}
\caption{The curve $\ds h(\rho)=\psi(1+\rho)\!+\!\gamma$
 passes through points $\ds h(n)=\sum_{k=1}^n\frac{1}{k}\;$.}
\end{center}
\end{figure}

\subsection{Series of the Riemann Zeta Function up to Finite Terms}

The analytic continuation (\ref{harmonic}) can be generalised.

The Riemann zeta function \cite{titchmarsh} is defined as
\be\zeta(s)\,=\,\sum_{k=1}^{\infty}\frac{1}{k^s}\quad,\quad \re(s)>1\ee

The series of the Riemann zeta function up to finite terms is 
defined as the Stegan-Riemann zeta function \cite{math_defn}
\bea\zeta(s,n)&=&\sum_{k=1}^{n}\frac{1}{k^s}\;=\;
\zeta(s) -\!\!\sum_{k=n+1}^{\infty}\frac{1}{k^s}\quad,\quad\re(s)>1\\
&=&\frac{d^n}{dx^n}\,\log(\Gamma(1+\hat{s}))\,
\bigg|_{\hat{s}=0}^{\hat{s}=s}\eea
Its analytic continuation is then given by simply replacing 
$\ds\frac{d^n}{dx^n}$ with the operator $D^w,\,w\in\c\,$,
\be\zeta(s,w)\;\;=\;\;D^w_{\hat{s}}\,\log(\Gamma(1+\hat{s}))\,
\Big|_{\hat{s}=0}^{\hat{s}=s}\label{D_log}
\qquad\qquad\qquad\qquad\quad\ee
which can be evaluated when $\log(\Gamma(1\!+\!\hat{s}))$ is expressed in
the form of asymptotic series \cite{math_defn}.

\subsection{Tables of Analytic continued Integrals}

An interesting consequence is that perhaps new editions of Tables of
Integrals may have to be compiled, eg. compute the coefficient of
$x^k$ term, $w(\rho,r,a,k)$, in the evaluation of
$$\ds \int x^r (\log x)^{\!a} (dx)^\rho$$

\section{Application: Calculus $-$\\
{\normalsize\bf Non-integer Power Series,\\
Breaking of Leibniz rule and Chain rule}\label{sec_nonint_pow_series}}

\subsection{Non-integer Power Series}

In {\em Type I}$\;$ Fractional Calculus \cite{frac_calculus},
we would write the power series for $\exp(x)$ as
\be \exp(x)\,=\,\lim_{\epsilon\to 0}\sum_{n=-\infty}^{\infty}
\frac{1}{\Gamma(1+\epsilon+n)}\,x^{n+\epsilon}\,=\,
\sum_{n\in\z}\frac{1}{\Gamma(1+n)}\,x^n\ee

\noindent In {\em Type II}$\,$, we would write it simply as
\be \exp(x)\,=\,\sum_{n=0}^{\infty}\frac{1}{\Gamma(1+n)}\,x^n\ee
since $D^\sigma_{\!x} x^r\,=\,0\quad\mbox{ for }\;\sigma>r\,,\,r\ge0$.

Thus, in {\em Type II}$\,$, operator $D$ acting on a power series real
$\sigma$-times
\be D^\sigma_{\!x} \sum_{k=0}^{\infty} a_k x^k = 
\!\!\sum_{k\,=\left\lceil\big.\sigma\right\rceil}^{\infty} 
\!a_k\,\frac{\Gamma(1\!+\!k)}{\Gamma(1\!+\!k\!-\!\sigma)}\,x^{k-\sigma} 
\quad\mbox{for}\;\;\sigma>0 \label{nonint_pow_series}\ee
where $\left\lceil\Big.\;\;\right\rceil$ denotes taking the integer ceiling.

When $\sigma$ is not an integer, r.h.s. of (\ref{nonint_pow_series})
is a non-integer power series.

Define the notation
\be f(\sigma,x) \equiv D^\sigma_{\!x}f(x)\ee
Think of $\sigma$ in the following way: the one-variable function $f(x)$ is
extended to a two-variable function $f(\sigma,x)$ in which $\sigma$
has now become a variable of the extended function.
\be\ba{l}\cos(\sigma,x)\,=\,D^\sigma_{\!x}\cos(x)\,=\, 
\ds D^\sigma_{\!x}\sum_{k=0}^{\infty}\frac{(-1)^k}{(2k)!}\,x^{2k}
\\\\
\quad=\ds\sum_{k\,=\,W\!c(\sigma)}^{\infty}\!
\frac{(-1)^k}{(2k)!}\,\frac{\Gamma(1\!+\!2k)}{\Gamma(1\!+\!2k\!-\!\sigma)}\,
x^{2k-\sigma}\qquad\quad\;\;\mbox{for}\quad\sigma>0\ea\ee
\be \mbox{where }\qquad
W\!c(\sigma)=\left\lceil\frac{\sigma}{2}+1\right\rceil-1
\qquad\qquad\qquad\qquad\ee

\be\ba{l}\sin(\sigma,x)\,=\,D^\sigma_{\!x}\sin(x)\,=\, 
\ds
D^\sigma_{\!x}\sum_{k=0}^{\infty}\frac{(-1)^k}{(2k\!+\!1)!}\,x^{2k+1}
\\\\
\quad=\ds\sum_{k\,=\,W\!s(\sigma)}^{\infty}\!
\frac{(-1)^k}{(2k\!+\!1)!}\,\frac{\Gamma(2(k\!+\!1))}
{\Gamma(2(k\!+\!1)\!-\!\sigma)}\,
x^{2k+1-\sigma}\quad\mbox{for}\quad \sigma>0\ea\ee
\be \mbox{where }\qquad
W\!s(\sigma)=\left\lceil\frac{\sigma}{2}+\frac{1}{2}\right\rceil-1
\qquad\qquad\qquad\qquad\ee

\be\ba{l}\exp(\sigma,x)\,=\,D^\sigma_{\!x}\exp(x)\,=\, 
\ds
D^\sigma_{\!x}\sum_{k=0}^{\infty}\frac{(-1)^k}{k!}\,x^k
\\\\
\quad=\ds\sum_{k\,=\,\left\lceil\big.\sigma\right\rceil}^{\infty}\!
\frac{(-1)^k}{k!}\,\frac{\Gamma(1\!+\!k)}
{\Gamma(1\!+\!k\!-\!\sigma)}\,
x^{k-\sigma}\qquad\qquad\quad\;\;\mbox{for}\quad \sigma>0\ea\ee
\begin{table}[hbt]
\caption{Some tabulated values of $W\!c(\sigma)$ and $W\!s(\sigma)$}
$$\begin{array}{|c||c|c|c|c|c|c|c|c|c|c|c|}
\hline \sigma         &0.0&0.5&1.0&1.5&2.0&2.5&3.0&3.5&4.0&4.5&\cdots\\
\hline W\!c(\sigma)&0  &1  &1  &1  &1  &2  &2  &2  &2  &3&\cdots\\
\hline W\!s(\sigma)&0  &0  &0  &1  &1  &1  &1  &2  &2  &2&\cdots\\
\hline \hline\end{array}$$
\label{W_table}
\end{table}

We then find that 
\be\quad\ba{c}\cos(\sigma,0)\,=\,
\left\{\ba{l} 1\;,\quad\mbox{if }\;\sigma\in 2\,\z\\
0 \;,\quad\mbox{otherwise}\ea\right.\quad,\quad 
\sin(\sigma,0)\,=\,
\left\{\ba{l} 1\;,\quad\mbox{if }\;\sigma\in 2\,\z\!+\!1\\
0 \;,\quad\mbox{otherwise}\ea\right.\\
\exp(\sigma,0)\,=\,
\left\{\ba{l} 1\;,\quad\mbox{if }\;\sigma\in\z\\
0 \;,\quad\mbox{otherwise}\ea\right.\ea\ee
$$\ba{rclcl}
\cos(\sigma,x)&=&\pm\,\sin(\sigma\!\pm\!1,x)&=&-\cos(\sigma\!\pm\!2,x)\\
\sin(\sigma,x)&=&\mp\cos(\sigma\!\pm\!1,x)&=&-\,\sin(\sigma\!\pm\!2,x)\\
\exp(\sigma,x)&=&\;\;\;\exp(\sigma\!\pm\!1,x)\ea$$
in agreement with the definitions of $\,\cos(x),\;\sin(x)$ and
$\exp(x)\,$ when $\,\sigma\in\z\,$.

\begin{figure}[hbt]
\begin{center}
\begin{tabular}{llll}
$\cos(0.01,\!x)$&$\cos(0.25,\!x)$&$\cos(0.5,\!x)$&$\cos(0.75\!,x)$\\
\psfig{figure=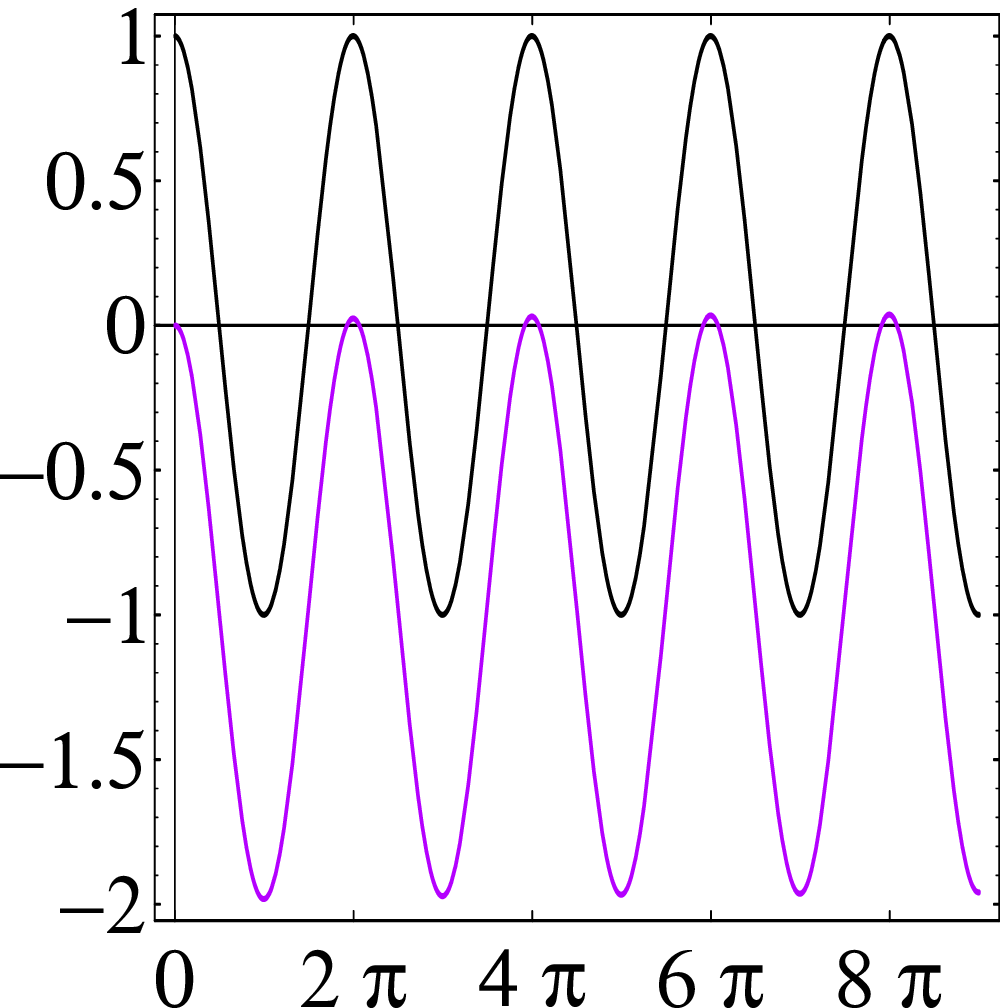,height=80pt}&
\psfig{figure=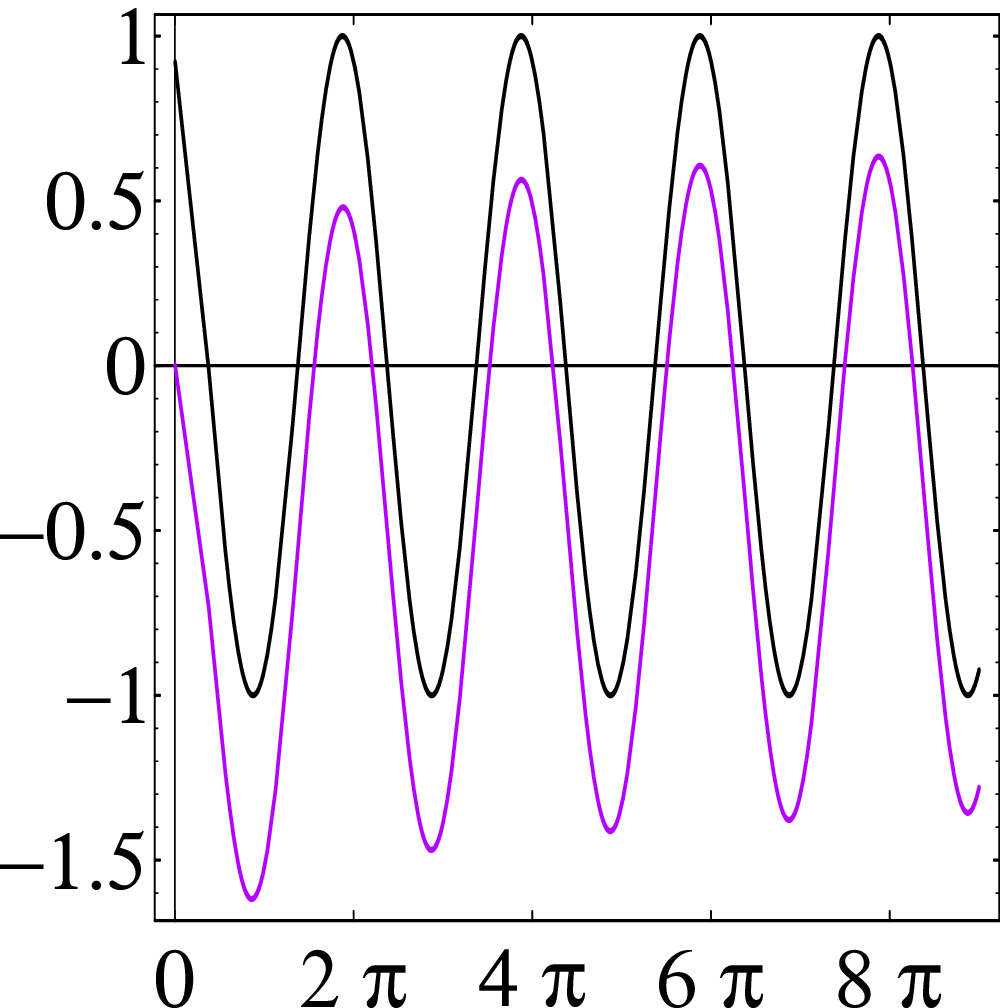,height=80pt}&
\psfig{figure=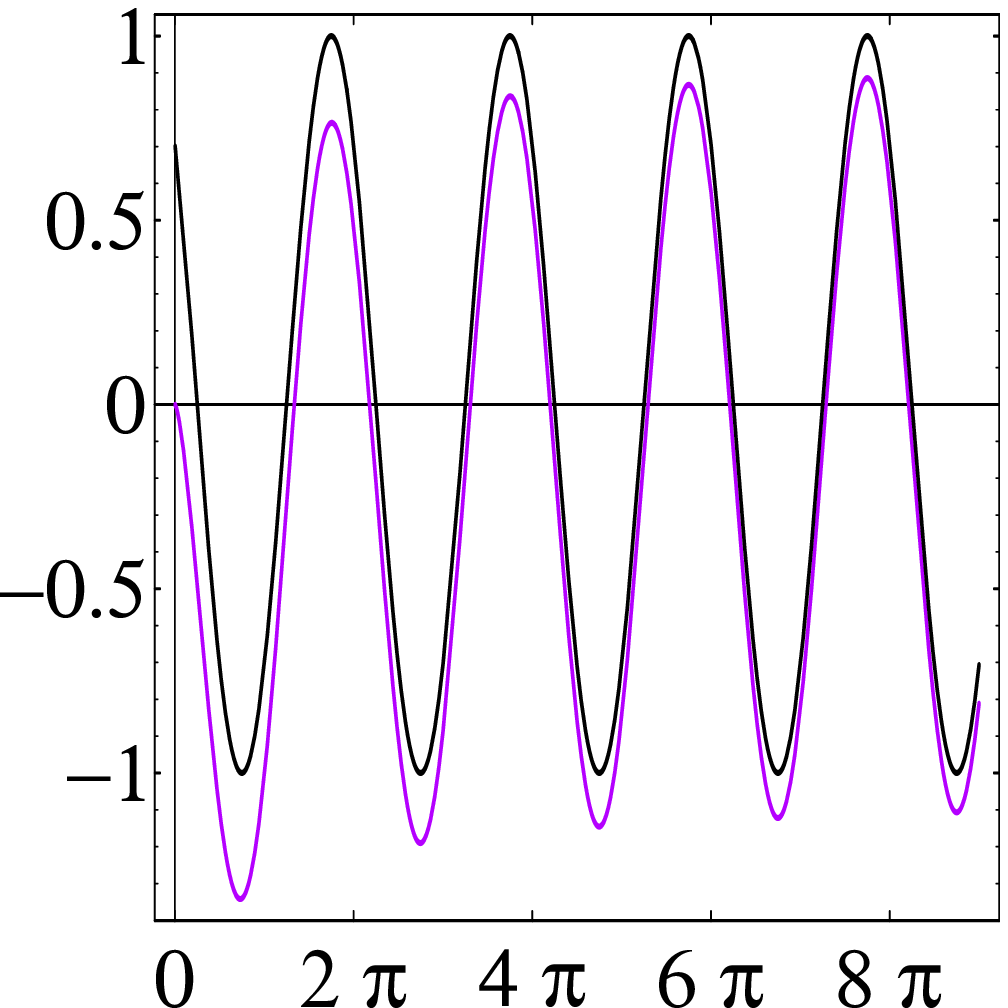,height=80pt}&
\psfig{figure=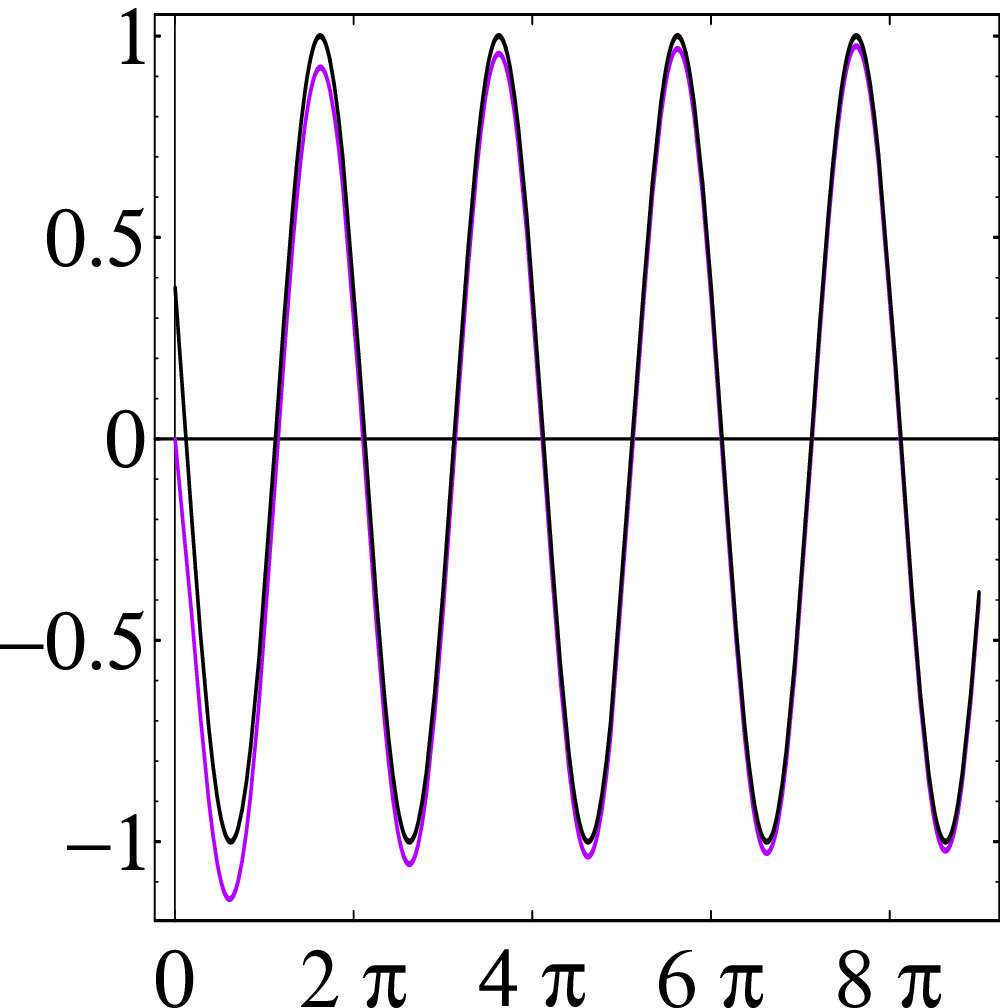,height=80pt}\\
$\qquad\qquad x$&$\qquad\qquad x$&$\qquad\qquad x$&$\qquad\qquad x$\\
\\
$\cos(1.25,\!x)$&$\cos(1.5,\!x)$&$\cos(1.99,\!x)$&$\cos(2.01,\!x)$\\
\psfig{figure=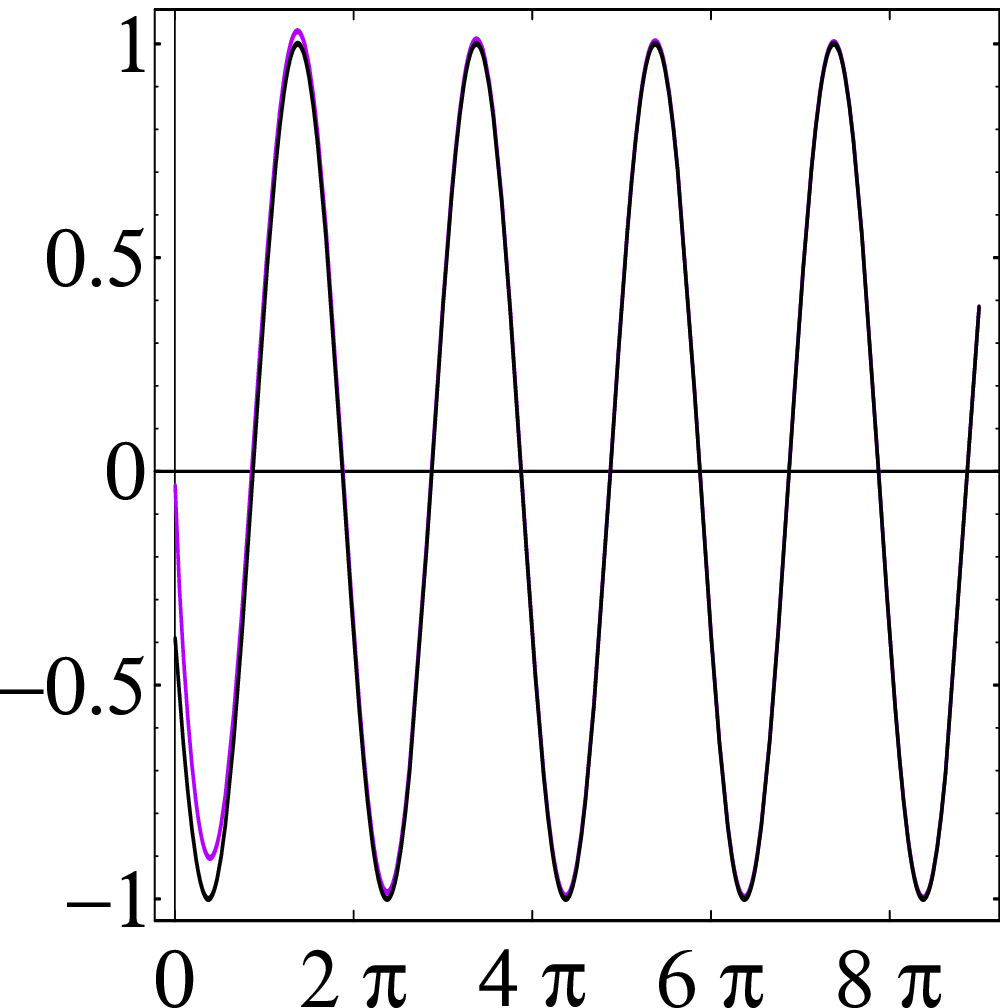,height=80pt}&
\psfig{figure=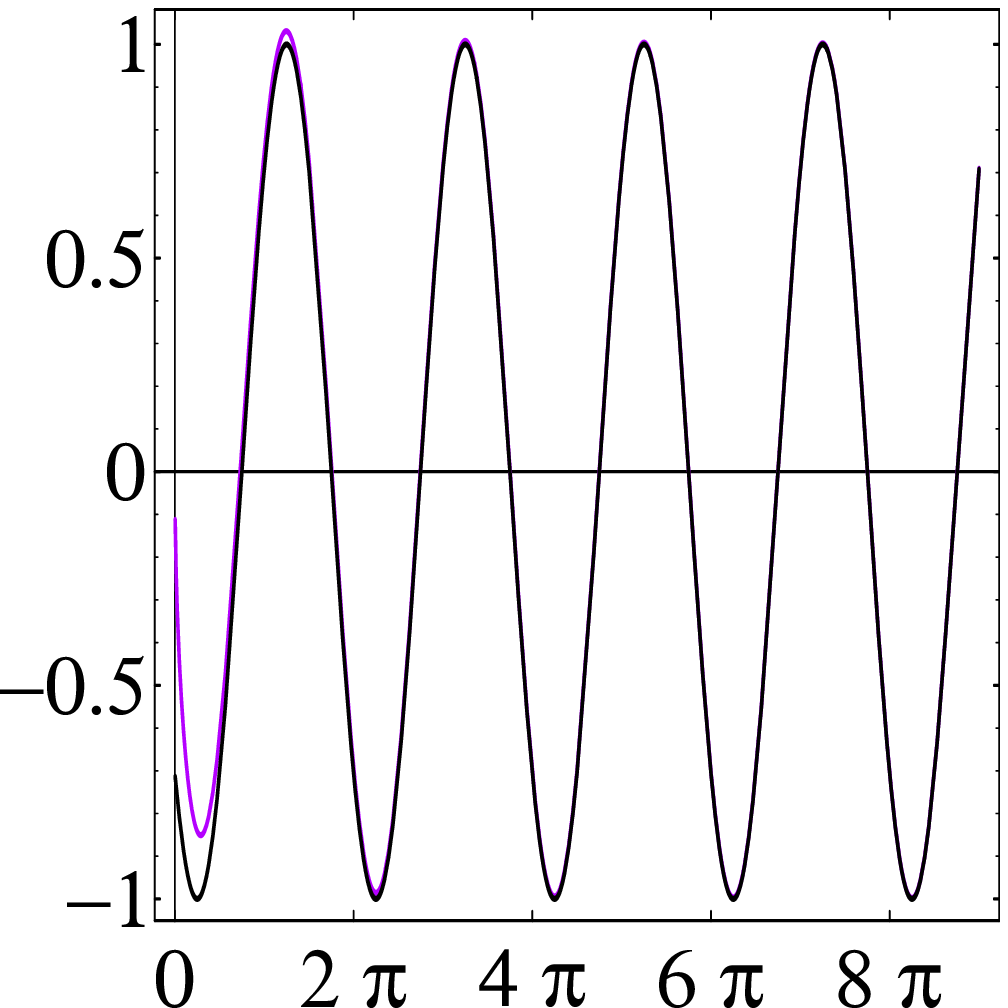,height=80pt}&
\psfig{figure=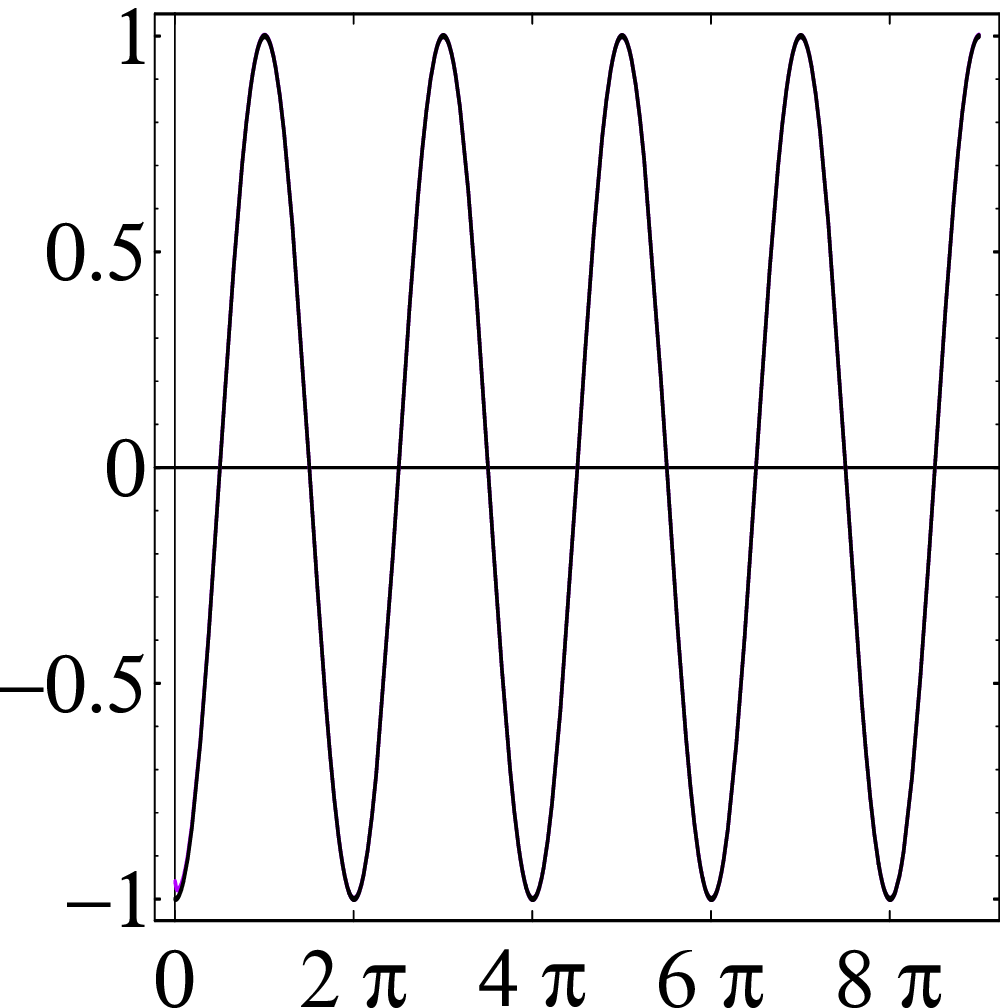,height=80pt}&
\psfig{figure=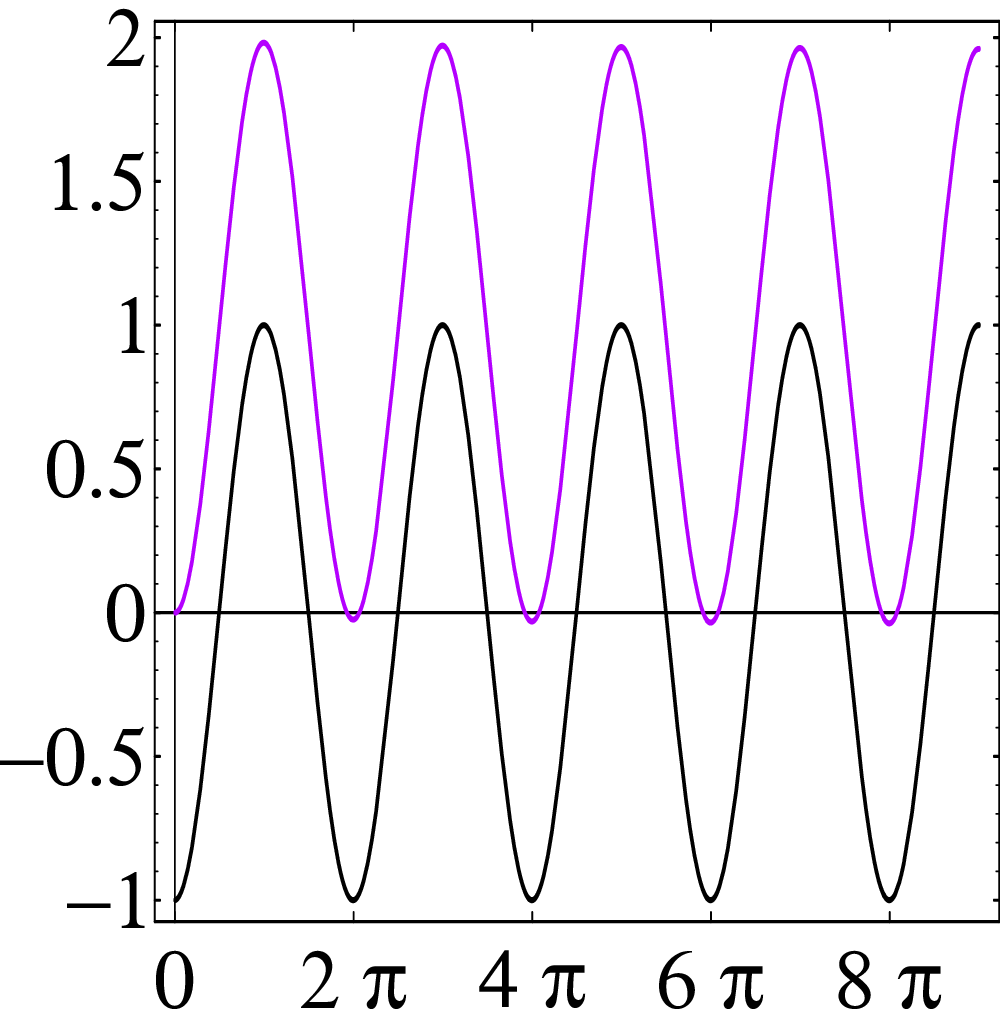,height=80pt}\\
$\qquad\qquad x$&$\qquad\qquad x$&$\qquad\qquad x$&$\qquad\qquad x$
\end{tabular}
\caption{Asymptotic limit: $\cos(\sigma,x)\equiv
D^\sigma_{\!x}\cos(x)\sim\cos(x\!+\!\ds\frac{\pi}{2}\sigma)$ as 
$x\to\infty$. \label{cos_asymp_lim}}
\end{center}
\end{figure}

\begin{figure}[hbt]
\begin{center}
\begin{tabular}{c}
$\left(\Big.\exp(\sigma,x) - \exp(x)\right)\qquad\qquad\qquad$\\
\psfig{figure=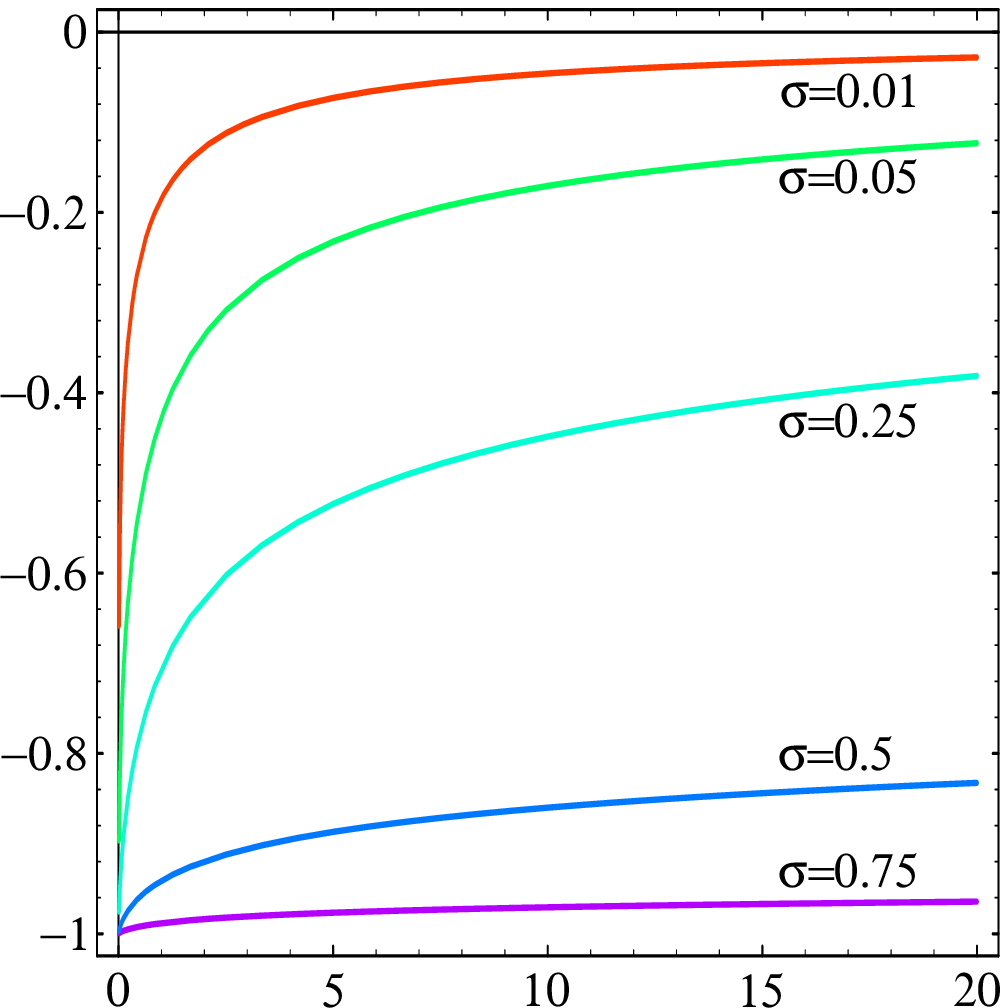,height=140pt}\\
$\qquad\quad x$
\end{tabular}
\caption{Asymptotic limit: $\exp(\sigma,x) - \exp(x) \to 0\;$ as 
$\,x\to\infty$.
\label{exp_asymp_lim}}
\end{center}
\end{figure}

In addition, it can be observed from Figures \ref{cos_asymp_lim} and
\ref{exp_asymp_lim} that there exist asymptotic limits
\be\left.\ba{rcl} \Bigg.\cos(\sigma,x)&\!\!\sim\!&
  \cos(x\!+\!\ds\frac{\pi}{2}\,\sigma)\\
  \Bigg.\sin(\sigma,x)&\!\!\sim\!&\,
  \sin(x\!+\!\ds\frac{\pi}{2}\,\sigma)\\
  \Bigg.\exp(\sigma,x)&\!\!\sim\!&
  \exp(x)
\ea\right\}\;\forall\;\sigma\;\mbox{ as }\;x\to\infty\ee
which remain to be proved analytically.

In fact, any function with a power series definition, eg. Bessel
functions, Fourier series, etc., can similarly have a non-integer
power series generalisation. In addition, the generalisation of
$\,\cos(x)\,$ to $\,\cos(\sigma,x)\,$ can be further extended to the
case of complex $s$, eg. $\,\cos(s,x) \equiv D^s_{\!x}\,\cos(x)$.

\subsection{Breaking of Leibniz rule and Chain rule\label{sec_breaking_rules}}

By definition, $D^n$ with integer $n$ obeys Leibniz rule
\be D^n_{\!x}\left\{f(x)\,g(x)\right\}\,=\,\sum_{k=0}^n {n\choose k}\!
\left(D^{n-k}_{\!x} f(x)\right)\left(D^k_{\!x} g(x)\right)\ee
and $D^1$ obeys Chain rule
\be D^1_{\!x}\,g\!\left(\Big.f(x)\right) = \left( D^1_{\!f}\,g\right)
\left( D^1_{\!x}f\right)\ee
but $D^s$ with complex $s$ does not in general.

However, by observing that
$$D^1_{\!x}\{fg\}\,=\,(D^1_{\!x}f)\,D^1_{\!f}\{fg\}\,+\,
(D^1_{\!x}\,g)\,D^1_{\!g}\{fg\}\,=\,(D^1_{\!x}f)\,g\,+\,
(D^1_{\!x}\,g)f$$
$$\Rightarrow\quad D^1_{\!x}\,=\,(D^1_{\!x}f)\,D^1_{\!f}\,+\,
(D^1_{\!x}\,g)\,D^1_{\!g}\qquad\qquad\qquad\qquad\qquad\qquad$$
we can express $D^s$ where ${s\in\c}\,$ in terms of nested sums of $D^1$'s 
which we can evaluate.
\bea D^s_{\!x} &=&
\left(\Big.(D^1_{\!x}f)\,D^1_{\!f}+(D^1_{\!x}\,g)\,
D^1_{\!g}\right)^{\!s} \nn\\
&=& \left(\bigg.1-\left(1-\left(\Big.(D^1_{\!x}f)\,D^1_{\!f}+
(D^1_{\!x}\,g)\,D^1_{\!g}\right)\right)\right)^{\!s} \nn\\
&=& 1 + \sum_{k=1}^\infty \frac{(-1)^k}{k!}\!\left[\prod_{m=1}^{k}
(s\!-\!m\!+\!1)\right]\!  \left(\Big.(D^1_{\!x}
f)\,D^1_{\!f}+(D^1_{\!x}\,g)\,D^1_{\!g}\right)^{\!k}
\label{D_series}\eea

Now, we can evaluate expressions of the form of 
$\,D^s_{\!x}\left\{f(x)\,g(x)\right\}\,$ and $\,D^s_{\!x}\,
g\!\left(\Big.f(x)\right)\,$ by simply by substituting $D^s$ with the
series on the r.h.s. of (\ref{D_series}).  See Section
\ref{sec_op_series}, equation (\ref{op_series}) for the problem of
convergence of the series.

\section{Application: Group Theory $-$\\
\\
\normalsize{\bf Analytic continuation of Groups:
R}${}_{\mbox{(mod {\it n})}}\;${\bf Groups,\\
Pseudo-Groups, and Symmetry Breaking/Deforming in Groups}
\label{sec_group}}

\subsection{R${}_{\mbox{\rm (mod {\it n})}}$ Groups}

The differential operator, and its inverse --- integral operator, can
act on different functional spaces to generate different {\em discrete}
groups.  These are groups of operators, ie. groups with operators as
elements.

$$\fbox{$\ds\quad \frac{d^n}{dx^n} f(x) = f(x) \quad $}$$
$$\ba{|l|l|l|}
\hline
\mbox{Order} & \;\,\mbox{Functional Space} & 
\quad\quad\quad\quad\mbox{Symmetry Group} \\
\hline \hline
n = 1 & f(x) = \;\exp(x) & \ds \z_1  = \left\{ {\bf id} \right\}, 
  \;\frac{d}{dx} \equiv {\bf id} = \frac{d^0}{dx^0} \\
n = 2 & f(x) = \left\{ \ba{c}\!\!\cosh(x)\\ \!\!\sinh(x)\ea \right. 
 & \ds \z_2 = \left\{ {\bf id} \,, \frac{d}{dx} \right\}, 
  \;\frac{d^2}{dx^2} \equiv {\bf id}\\
n = 4 & f(x) = \left\{ \ba{c}\!\!\pm\cos(x)\\ \!\!\pm\sin(x)\ea \right. 
 & \ds \z_4 = \left\{ {\bf id} \,, \frac{d}{dx} \,, \frac{d^2}{dx^2} \,,
   \frac{d^3}{dx^3} \right\}, \;\frac{d^4}{dx^4} \equiv {\bf id}\\
\hline \ea $$
\begin{figure}[hbt]
\begin{center}
\begin{tabular}{c}
\psfig{figure=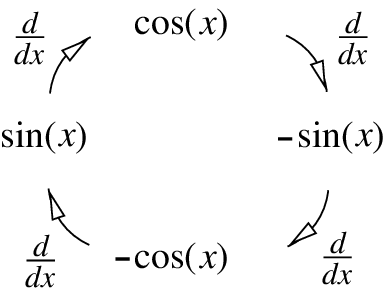}
\end{tabular}
\caption{$\z_4$ group flow diagram of}
\begin{tabular}{c}
$\ds\frac{d}{dx}$ acting on functional space $\left\{\pm\cos(x),
  \pm\sin(x)\right\}$.
\end{tabular}
\end{center}
\end{figure}

If elements of the functional space are extended from functions $f(x)$
to their analytic continuations $f(\sigma,x) = D^\sigma f(x)$ with
real $\sigma$, operators $D^{\sigma'}$ acting on these extended
functional spaces will generate {\em continuous} groups or {\em Lie}
groups, eg.,

$D^{\sigma'}_{\!x}$ acting on the functional space $\left\{
\cos(\sigma,x) \;\Big|\; \sigma \in [0,4) \,,\, x \in [0,\infty) 
\right\}$ generates a natural analytic continuation of the
$\mbox{\bf Z}_4$ group,
$$\left\{ D^{\sigma'}_{\!x} \;\Big|\; \sigma' \in [0,4) \right\} \,,\,
D^4_{\!x} \equiv \mbox{\bf id}$$
By analogy to the concept of (mod $n$) congruence in Number Theory, we
denote this analytic continued group $\r_{\mbox{(\footnotesize mod
    4)}}$.

$$\fbox{$\quad D^{\sigma'}_{\!x} f(\sigma,x) = f(\sigma,x) \quad $}$$
$$\ba{|l|l|l|l|}
\hline
\mbox{Order} & \quad\;\mbox{Functional Space} & 
\quad\quad\quad\quad\mbox{Symmetry Group} \\
\hline \hline
\sigma' = 1 & f(\sigma,x) = \;\exp(\sigma,x) &
  \r_{\mbox{(\footnotesize mod 1)}} = \left\{ D^{\sigma'}_{\!x} \;\Big|\;
  \sigma' \in [0,1) \right\}, D^1_{\!x} \equiv {\bf id} \\
\sigma' = 2 & f(\sigma,x) = \left\{ \ba{c}\!\!\cosh(\sigma,x)\\
  \!\!\sinh(\sigma,x)\ea \right. & \r_{\mbox{(\footnotesize mod 2)}}
  = \left\{ D^{\sigma'}_{\!x} \;\Big|\; \sigma' \in [0,2) \right\},
  D^2_{\!x} \equiv {\bf id} \\
\sigma' = 4 & f(\sigma,x) = \left\{ \ba{c}\!\!\pm\cos(\sigma,x)\\
  \!\!\pm\sin(\sigma,x)\ea \right. & \r_{\mbox{(\footnotesize mod 4)}}
  = \left\{ D^{\sigma'}_{\!x} \;\Big|\; \sigma' \in [0,4) \right\},
  D^4_{\!x} \equiv {\bf id} \\
\hline \ea $$
\begin{figure}[hbt]
\begin{center}
\begin{tabular}{c}
\psfig{figure=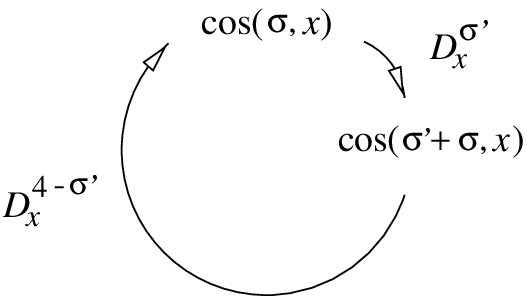}
\end{tabular}
\caption{$\r_{\mbox{(\footnotesize mod 4)}}$ group flow diagram of}
\begin{tabular}{c}
$D^{\sigma'}_{\!x}$ acting on functional space $\left\{
\cos(\sigma,x) \;\Big|\; \sigma \in [0,4)\right\}$.
\end{tabular}
\end{center}
\end{figure}

For complex $s=\sigma\!+\!it\,,\,s'=\sigma'\!+\!it'$,\\
$D^{s'}$ acting on functional space $\left\{\cos(s,x) \;\Big|\; \sigma
  \in [0,4) \,,\, t \in \r \,,\, x \in [0,\infty) \right\}$\\
generates a Lie group $\,\r_{\mbox{(\footnotesize mod 4)}}\,{\sf X}\,\r\,$
since $D^{it'}$ commutes with $D^{\sigma'}$ and so $D^{it'}$ acts
independently from $D^{\sigma'}$.

\begin{figure}[hbt]
\begin{center}
\begin{tabular}{c}
\psfig{figure=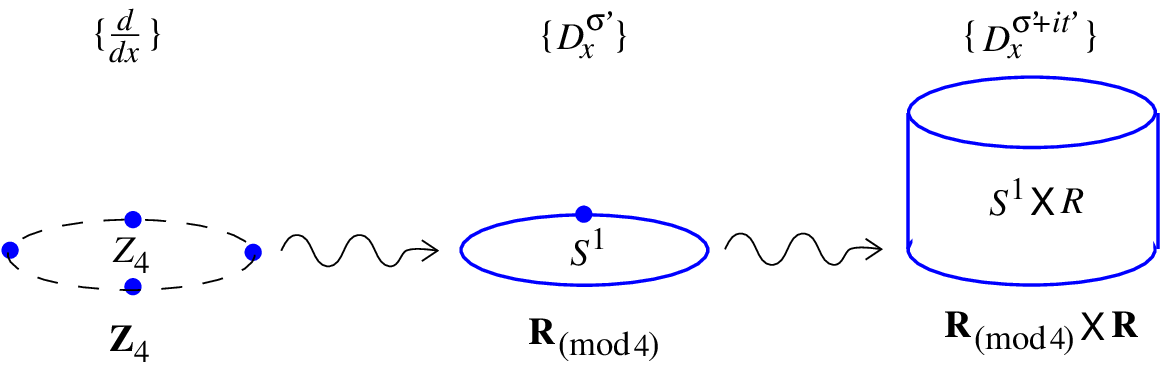}
\end{tabular}
\begin{tabular}{ccc}
Group & & Topology\\
$\overbrace{\z_4 \to \r_{\mbox{(\footnotesize mod 4)}} \to
 \r_{\mbox{(\footnotesize mod 4)}}\,{\sf X}\,\r}$
&$\;\;\Rightarrow\;\;$&
$\overbrace{Z_4 \,\to\, S^1 \,\to\, S^1\,{\sf  X}\,R}$
\end{tabular}
\caption{Topology change of groups.\label{D_group_topo}}
\end{center}
\end{figure}

In general, the topology of such analytic continued groups progresses
from sets of points on a circle $S^1$ for discrete groups
generated by $d/dx$, to a circle $S^1$ for Lie groups
generated by $D^{\sigma'}_{\!x}$, and to a 2-dimensional cylinder
$\,S^1${\sf X}$R\,$ for Lie group generated by $D^{\sigma'+it'}_{\!x}$ as
illustrated in Figure \ref{D_group_topo}.

\subsection{Pseudo-Groups in {\em Type II}$\;$ Fractional Calculus}

Consider the analytic continuation of the group elements of $SO(2)$
(the group of rotation in a plane) in {\em Type II}$\;$ Fractional
Calculus,
\be\left.\ba{c}R(\theta) = \left( \ba{rl}
 \cos\,\theta & \sin\,\theta\\
-\sin\,\theta & \cos\,\theta
\ea\right)\\
\\\mbox{where }\theta\in [0,2\,\pi)\ea\right\}
\longmapsto
\left\{\ba{c}R(\sigma,\theta) = \left( \ba{rl}
 \cos(\sigma,\theta) & \sin(\sigma,\theta)\\
-\sin(\sigma,\theta) & \cos(\sigma,\theta)
\ea\right)\\
\\\mbox{where }\theta\in[0,\infty)\ea\right.\ee
$R(\sigma,\theta)$ forms a set of sets, paramatrized at 2 levels.  The
set of sets is parametrized by $\sigma$, and each of these sets is
further parametrized by $\theta$. Denote the set of these sets as
$SO(2;\sigma,\theta)$.

Since\\
\be R(0,\theta) \in SO(2)\;\;\forall\,\theta\;\quad\mbox{and}\quad
R(\sigma,\theta)\sim R(0,\theta\!+\!\frac{\pi}{2}\,\sigma)\;\mbox{ as }\;
\theta \to \infty\ee
we are motivated to introduce the concept of pseudo-groups.

A {\em pseudo-group} $\,G(\rho_1,\rho_2,\dots,\rho_k)\,$ of a group
$\,G\,$ is a set which gradually acquires the group properties or
satisfies the group axioms of $\,G\,$ as {\em some} of the parameters
$\,\rho_1,\rho_2,\dots,\rho_k\,$ of the set approach limiting values
or tend asymptotically to infinity.

$SO(2;\sigma,\theta)$ is a pseudo-group of $SO(2)$ since it is
isomorphic to $SO(2)$ when
\begin{itemize}
\item the parameter ${\sigma\to n\in\z}$ while $\,\theta\,$
  varies freely in the interval $[\,0,\infty)$,
$$R(\sigma,\theta)\,.\,R(\sigma',\theta)\,\sim\,
  R(\sigma\!+\!\sigma',\theta)\quad\mbox{as}\;\;\sigma,\sigma'\to n,n'\in\z$$
  \be\Rightarrow\quad\lim_{\sigma\to n\in\z} SO(2;\sigma,\theta)
  \,\cong\,SO(2)\ee
\item (alternatively) the parameter ${\theta\to\infty}$ while
  $\sigma$ varies freely in the interval $(0,2)$,
  $$R(\sigma,\theta)\,.\,R(\sigma,\theta')\,\sim\,
  R(0,\theta\!+\!\ds\frac{\pi}{2}\sigma)\,.\,
  R(0,\theta'\!\!+\!\ds\frac{\pi}{2}\sigma)\,=\,
  R(0,\theta\!+\!\theta'\!\!+\!\pi\sigma)$$
  $$\quad\mbox{as}\;\;\theta,\theta'\to\infty$$
  \be\Rightarrow\quad \lim_{\theta\to\infty} SO(2;\sigma,\theta)
  \,\cong\,SO(2)\ee
\end{itemize}
\begin{figure}[hbt]
\begin{center}
\begin{tabular}{c}
\psfig{figure=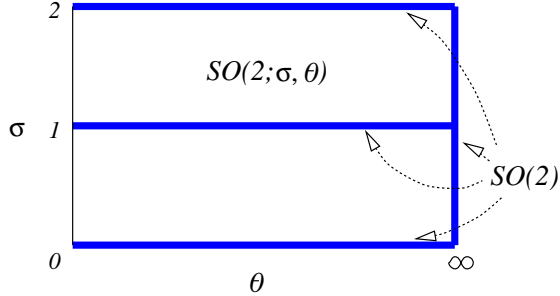}
\end{tabular}
\caption{$SO(2;\sigma,\theta)$ plane diagram.}
\end{center}
\end{figure}

We can define a group property deviation measure ${\cal
  W}(G(\sigma,x),G\bigm|\sigma,x)$ for a pseudo-group $G(\sigma,x)$, a
measure of how much group property the pseudo-group has lost or
deviated from the associated ``parent'' group $G$ from which it is
analytic continued. When the pseudo-group becomes isomorphic to the
parent group for certain values of the parameter, the measure should
be zero.

\begin{figure}[hbt]
\begin{center}
\begin{tabular}{c}
\psfig{figure=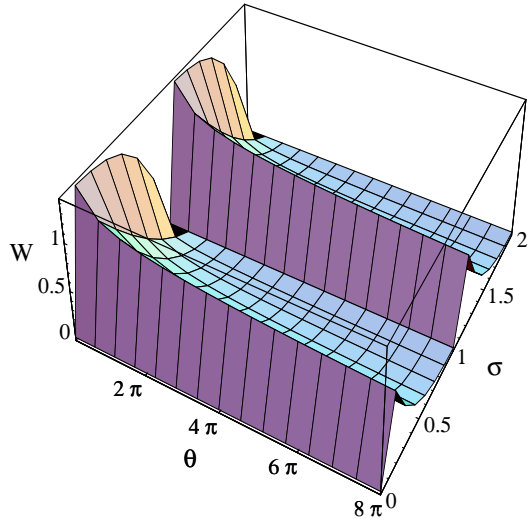,height=200pt}
\end{tabular}
\caption{Measure ${\cal W}$ of $SO(2;\sigma,\theta)$.\label{SO2_measure}}
\end{center}
\end{figure}

For the case of pseudo-group $SO(2;\sigma,\theta)$,
\bea\lefteqn{{\cal W}(SO(2;\sigma,\theta),SO(2)\bigm|\sigma,\theta)}\nn\\
&=&\left|\,\Big.R(\sigma,\theta) - R(\theta\!+\!\frac{\pi}{2}\sigma)\;
\right|\nn\\
&=&\left|\ba{cc}\;\;\cos(\sigma,\theta)-\cos(\theta
\!+\!\ds\frac{\pi}{2}\sigma)&
\;\sin(\sigma,\theta)-\sin(\theta
\!+\!\ds\frac{\pi}{2}\sigma)\nn\\
-\!\left(\sin(\sigma,\theta)-\sin(\theta
\!+\!\ds\frac{\pi}{2}\sigma)\right)&
\;\cos(\sigma,\theta)-\cos(\theta
\!+\!\ds\frac{\pi}{2}\sigma)\ea\right|\\
&=&\left|\ba{cc}\;\;\delta\!\cos&\;\;\delta\!\sin\\
-\delta\!\sin&\;\;\delta\!\cos\ea\right|\nn\\
&=&\Bigg.\sqrt{\left(\delta\!\cos+\delta\!\sin\right)^2+
\left(-\delta\!\sin+\delta\!\cos\right)^2}\nn\\
&=&\Bigg.\sqrt{2\left(\delta\!\cos^{\,2}+\delta\!\sin^2\right)}\nn\\
&=&\sqrt{2\left(\left(\cos(\sigma,\theta)-\cos(\theta
\!+\!\ds\frac{\pi}{2}\sigma)\right)^2+\left(\sin(\sigma,
\theta)-\sin(\theta\!+\!\ds\frac{\pi}{2}\sigma)\right)^2\right)}
\eea
satisfies the requirement. See Figure \ref{SO2_measure}.

Similarly for the simple case of $U(1;\sigma,i x)$, a pseudo-group of
$U(1)$ where $x\in\r$,
$$\exp(\sigma,i x_1)\;\exp(\sigma',i x_2)\,\sim\,
\exp(\sigma,i(x_1\!+\!x_2))\quad\mbox{as}\;\;\sigma,\sigma'\to n,n'\in\z$$
$$\Rightarrow\quad\lim_{\sigma\to n\in\z} U(1;\sigma,x)\,\cong\,U(1)$$
$$\exp(\sigma,i x_1)\;\exp(\sigma,i x_2 )\,\sim\,\exp(\sigma,i(x_1\!+\!x_2))
\quad\mbox{as}\;\;x_1 ,x_2 \to \infty$$
$$\Rightarrow\quad \lim_{\theta\to\infty} U(1;\sigma,x)\,\cong\,U(1)$$
$${\cal W}(U(1;\sigma,ix),U(1)\bigm|\sigma,i x)\,=\,
\exp(\sigma,i x)-\exp(i x)$$
This measure was plotted in Figure \ref{exp_asymp_lim}.

\subsection{$SO(2;\sigma,\theta)$ Rotations and Deformations in {\em
    Type II}}

\begin{figure}[hbt]
\begin{center}
$\ba{l}\;\;\fbox{$\;\sigma\Big\backslash\theta\;$}\;\,\approx 
0\;\;\,\pi/16\;\;\,\pi/8\;\;\;\pi/4\;\;\;\pi/2\;\;\,3\pi/4\;\;\;\,\pi
\quad 3\pi/2\quad 2\pi\quad\;\, 4\pi\quad\;\, 6\pi \quad\;\, 8\pi\\
\ba{rl}
\raisebox{167pt}{$\ba{r}0\\\\0.001\\\\0.25\\\\0.5\\\\0.75
\\\\1\\\\1.25\\\\1.5\\\\1.75\\\\2\\\\3\\\\4 \ea$}&\!\!\!\!\!\!\!\!
\psfig{figure=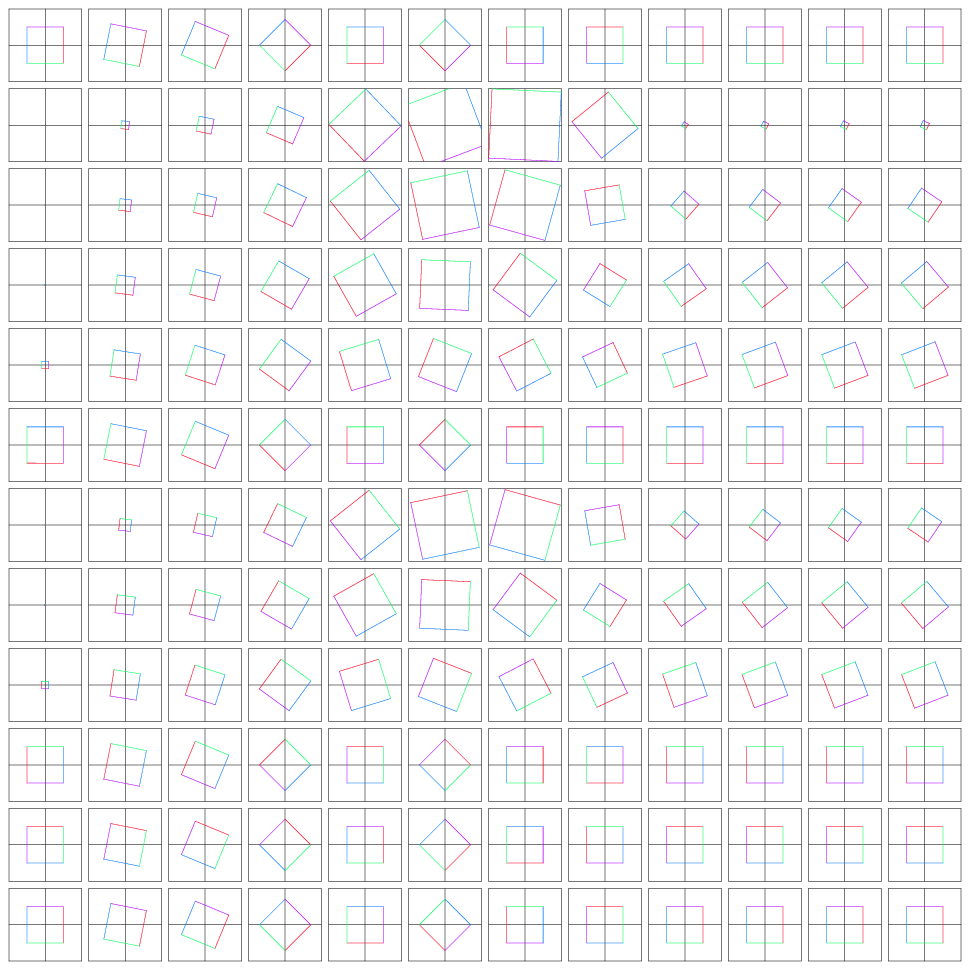,height=340pt}
\ea\ea$
\caption{$SO(2;\sigma,\theta)$ rotations and deformations 
(color illustration). \label{SO2_rot}}
\end{center}
\end{figure}

Figure \ref{SO2_rot} shows the effect of planar rotations and
deformations of $SO(2;\sigma,\theta)$ on a square with vertices
$\{\,(1,\sm1),\,(1,1),\,(\sm1,1),\,(\sm1,\sm1)\}$ on a sequence of
$(x,y)$ planes clipped by square windows of size
$x\!\in\![-2,2],\;y\!\in\![-2,2]$. The deformation effects seem to be
a combination of rotations and contractions/dilations.

\subsection{Symmetry Breaking/Deforming in Groups in {\em Type II}}

In the Higgs mechanism of Spontaneous Symmetry Breaking \cite{higgs},
\begin{itemize}
\item the symmetry of the effective potential $\,V_{e\!f\!f}\,$ in a
  Lagrangian density $\,\cal L\,$ with respect to a gauge group $\,G\,$ is
  preserved, while
  
\item the symmetry of the quantum state $\,\psi\,$ satisfying the
  equations of motion derived from $\,\cal L\,$ is broken and reduced
  to that of a subgroup, $\,H \subset G\,$.
\end{itemize}
The profile of $V_{e\!f\!f}$ changes with energy or temperature.  At
high energy or temperature, the symmetry of $\psi$ is restored from $\,H
\to G\,.$

In the case here, the symmetry breaking is very different.  The
symmetry in a group $\,G\,$ itself is broken to a subgroup $\,H
\subset G\,$ or ``deformed'' into an approximate symmetry of $\,G\,$.

Take $\,SO(3;(\sigma_1,x_1),(\sigma_2,x_2))$, a pseudo-group of
$\,SO(3)$, as an example. When both $\sigma_1,\sigma_2\,=\,0\,,\;
SO(3;(\sigma_1,x_1),(\sigma_2,x_2))$ is isomorphic to $SO(3)$.
Now, choose ${\;\sigma_1\not\in\z\;\,}$ and ${\;\sigma_2=0\,}$.\\
The $SO(3)$ symmetry is then $\left\{\ba{c}
  \mbox{broken}\\\!\!\mbox{approximate}\!\!\\\mbox{restored}\ea\right\}
\mbox{ for }\left\{\ba{c}
  \mbox{small}\\\!\!\mbox{intermediate}\!\!\\\mbox{large}\ea\right\}
\;x$, and the $SO(3)$ symmetry in a sphere is ``deformed'' to an approximate
$SO(3)$ symmetry or completely broken to $SO(2)$ in a plane depending
on the chosen values of $\,\sigma_1\,$ and $\,x$.  Now, set both
${\;\sigma_1,\sigma_2\not\in\z\,}$. The $SO(2)$ is further
``deformed'' to an approximate $SO(2)$ symmetry or broken to Identity.

The symmetry breaking/deforming sequence is then
$$SO(3;(\sigma_1,x_1),(\sigma_2,x_2))
\stackrel{\sigma_1,\,\sigma_2=\,0}{\to\!\!\!-\!\!\!-\!\!\!\longrightarrow}
SO(3) \stackrel{\sigma_1\not\in\,\z\,,\,\sigma_2=\,0}
{\to\!\!\!-\!\!\!-\!\!\!-\!\!\!-\!\!\!\longrightarrow}
SO(2) \stackrel{\sigma_1,\,\sigma_2\not\in\z}
{\to\!\!\!-\!\!\!-\!\!\!\longrightarrow}\,\id$$

Similarly for $SU(N;(\sigma_1,x_1),(\sigma_2,x_2),\cdots,(\sigma_N,x_N))$,
the symmetry breaking/deforming  sequence is
$$\ba{l}SU(N;(\sigma_1,x_1),(\sigma_2,x_2),\cdots,(\sigma_N,x_N))\\
\\
\ba{l}
\quad\stackrel{\sigma_1,\,\sigma_2,\cdots,\,\sigma_N=\,0}
{\hookrightarrow\!\!\!-\!\!\!-\!\!\!-\!\!\!-\!\!\!-\!\!\!-\!\!\!-\!\!\!
-\!\!\!-\!\!\!-\!\!\!-\!\!\!-\!\!\!-\!\!\!-\!\!\!-\!\!\!-\!\!\!-\!\!\!
\longrightarrow}\quad SU(N)\\
\qquad\stackrel{\sigma_1\not\in\,\z\,,\,\sigma_2,\cdots,\,\sigma_N=\,0}
{\hookrightarrow\!\!\!-\!\!\!-\!\!\!-\!\!\!-\!\!\!-\!\!\!-\!\!\!-\!\!\!
-\!\!\!-\!\!\!-\!\!\!-\!\!\!-\!\!\!-\!\!\!-\!\!\!-\!\!\!-\!\!\!-\!\!\!
\longrightarrow}\quad SU(N\!-\!1)\\
\qquad\qquad\qquad\quad\vdots\qquad\vdots\\
\qquad\qquad\stackrel{\sigma_1,\cdots,\,\sigma_{N-2}\not\in\,\z\,,\,
\sigma_{N-1},\,\sigma_N=\,0}
{\hookrightarrow\!\!\!-\!\!\!-\!\!\!-\!\!\!-\!\!\!-\!\!\!-\!\!\!-\!\!\!
-\!\!\!-\!\!\!-\!\!\!-\!\!\!-\!\!\!-\!\!\!-\!\!\!-\!\!\!-\!\!\!-\!\!\!
\longrightarrow}\quad SU(2)\\
\qquad\qquad\quad\stackrel{\sigma_1,\cdots,\,\sigma_{N-1}\not\in\,\z\,,\,
\sigma_N=\,0}
{\hookrightarrow\!\!\!-\!\!\!-\!\!\!-\!\!\!-\!\!\!-\!\!\!-\!\!\!-\!\!\!
-\!\!\!-\!\!\!-\!\!\!-\!\!\!-\!\!\!-\!\!\!-\!\!\!-\!\!\!-\!\!\!-\!\!\!
\longrightarrow}\quad U(1)\\
\qquad\qquad\qquad\stackrel{\sigma_1,\,\sigma_2,\cdots,\,\sigma_N\not\in\,\z}
{\hookrightarrow\!\!\!-\!\!\!-\!\!\!-\!\!\!-\!\!\!-\!\!\!-\!\!\!-\!\!\!
-\!\!\!-\!\!\!-\!\!\!-\!\!\!-\!\!\!-\!\!\!-\!\!\!-\!\!\!-\!\!\!-\!\!\!
\longrightarrow}\quad\id\ea\ea$$

Perhaps this mode of symmetry breaking/deforming in groups might have
some useful applications for models in Particle Physics and Cosmology.

In Particle Physics, the symmetry of the flavor of quarks are not
exact symmetry but only approximate symmetry of Gell-Mann's Eightfold
way $SU(3)$ \cite{SU(3)} or GUT $\,SU(5)$ \cite{SU(5)} because
different flavors of quarks have different masses.  Light quarks do
not transform exactly into heavy quarks, and perhaps the effects from
the presence of gluons and glueballs in composite particles need to be
added into the symmetry.  Perhaps $SU(3)$ and $SU(5)$ can be
``deformed'' in this way to an approximate symmetry that will fit the
phenomenological data better.

Now, take a pseudo-group to describe the product of residual exact and
approximate symmetries of present day Universe.  If we set $x\propto
T$, the temperature of the Universe, as we go back in time, the
temperature $T$ goes up, $x$ goes up, and we find that the approximate
and other broken symmetries are gradually being restored.  The rate
the symmetries are being restored will be dependent on the values of
$\sigma_1,\,\sigma_2,\cdots,\sigma_N$, the parameters of the
pseudo-group. The fully restored symmetry will be the symmetry of the
parent group of the pseudo-group. Qualitatively, this model resembles
the unification of gauge groups in Cosmology \cite{universe}.  It
might be interesting to study and develop this mode of symmetry
breaking/deforming for approximate symmetry groups (eg. iso-spin
group, Eightfold way SU(3), and GUT $\,SU(5)\,$) as well as for the
gauge groups.

\section{Application: Algebra $-$\\
{\normalsize\bf Analytic continuation of Dirac Equation and Algebra}
\label{sec_algebra}}

\begin{quote}{\em ``I think that there is a moral to this story,
    namely that it is more important to have beauty in one's equations
    than to have them fit experiment.''} --- P.A.M. Dirac in {\em
    Scientific American}, May (1963)\end{quote}

Dirac equation $\Rightarrow$ Klein-Gordon equation
$$\ba{l}\ds\Bigg. i\,\frac{\partial}{\partial t} \psi = \left(
-i\Big.\underline{\alpha}\,.\,\underline{\nabla} + \beta\,m
\right)\!\psi 
\quad\Rightarrow\quad\left(i\,\frac{\partial}{\partial
  t}\right)^{\!2}\!\psi = \left(
  -i\Big.\underline{\alpha}\,.\,\underline{\nabla} + \beta\,m
\right)^{\!2}\!\psi \\
\ds\Bigg.\Rightarrow\;\;-\,\frac{\partial^2}{\partial t^2}\psi = \left(
  -\sum_{i\,,\,j}\alpha_i\alpha_j\nabla_i\nabla_j 
  \,-\,i m\!\left(\Big.\underline{\alpha}\,\beta +
  \beta\,\underline{\alpha}\right)\!.\,\underline{\nabla} +
  \beta^2\,m^2 \right)\!\psi \\ 
\ds\Bigg.\Rightarrow\;\;-\,\frac{\partial^2}{\partial t^2}\psi
= \left( -\,\nabla^2\,+\, m^2\right)\!\psi 
\quad\mbox{where}\;\;\sum_{i_1,i_2,\dots,i_p}\!\!\!\equiv
\;\sum_{i_1}\sum_{i_2}\cdots\sum_{i_p}$$
\ea$$
giving Dirac Algebra \cite{dirac_eqn}
\be\Rightarrow\;\;\;\left\{\alpha_i\,,\,\alpha_j\right\} = 2
\,\delta_{ij}\,\id \quad,\quad \left\{\alpha_i\,,\,\beta\right\}
= \zero \quad,\quad \alpha^2_i = \beta^2 = \id\quad\;\;\;\ee

Think of $\ds \left\{i\,\frac{\partial}{\partial t}\,,\,
-i\,\underline{\nabla}\right\}$ as a basis. The basis can be analytic
continued with the $D^s$ operator.

Dirac equation can then be analytic continued to
\be e^{\pi\,i/p}\,D^{2/p}_t \psi = \left( \Big. 
e^{-\pi\,i/p}\,\underline{\alpha}^{(2/p)}.\, 
\underline{D}^{(2/p)} + \beta^{(2/p)}\,m^{2/p} \right)\!\psi 
\label{dirac_eqn_con}\ee
$\quad \ba{cl}\mbox{where}
 & p = 1 \,,\;\;(\ref{dirac_eqn_con})\;\Rightarrow\;\mbox{Klein-Gordon
   equation}\\
 & \qquad\qquad\qquad\qquad\alpha^{(2)}_i = \id \quad,\quad
   \beta^{(2)} = \id\\
 & p = 2 \,,\;\;(\ref{dirac_eqn_con})\;\Rightarrow\;\mbox{Dirac equation}\\
 & \qquad\qquad\qquad\qquad\alpha^{(1)}_i = \alpha_i \quad,\quad \beta^{(1)}
   = \beta
\ea$

Introduce the notation for generalised symmetrisation
\be\ds\left\{\Big. A_{i_1}\,,\,A_{i_2}\,,\dots,\,A_{i_p}\right\} =
\!\!\sum_{\perm(i_1,i_2,\dots,i_p)}\!\!\!\!\!\!
A_{\perm_1}A_{\perm_2}\dots A_{\perm_p}\ee
where the sum is over all permutations of the $p$ indices.

The generalised symmetrisation can be re-expressed in terms of a sum
of permutations of nested anti-commutators, eg.
\be\{a,b,c\} =
\frac{1}{2\,(1!)}\left(\Big.
  \{\{a,b\},c\}+\{\{b,c\},a\}+\{\{c,a\},b\}\right)\ee
\be\ba{l}\{a,b,c,d\}\;=\;\\
\ds\frac{1}{2\,(2!)}\left(\ba{l}
  {}\;\;\{\{\{a,b\},c\},d\}+\{\{\{b,c\},d\},a\}+
  \{\{\{c,d\},a\},b\}+\{\{\{d,a\},b\},c\}\!\!\\
  {}\!\!\!\!+\{\{\{a,b\},d\},c\}+\{\{\{b,c\},a\},d\}+
  \{\{\{c,d\},b\},a\}+\{\{\{d,a\},c\},b\}\!\!\\
  {}\!\!\!\!+\{\{\{a,c\},b\},d\}+\{\{\{a,c\},d\},b\}+
  \{\{\{b,d\},c\},a\}+\{\{\{b,d\},a\},c\}\!\!\\
  \ea\right)\ea\ee

Now, for $p = 3$,

{L.h.s.} :
$$\left(e^{\pi\,i/3}\,D^{2/3}_t\right)^{\!3}\!\psi
\,=\,e^{\pi\,i}\,D^2_t\psi\,=\,-D^2_t\psi\,=\,-\ds
\frac{\partial^2}{\partial t^2}\psi
\qquad\qquad\qquad\qquad\qquad$$

{R.h.s.} :
\vskip .1truein
$\;\left(e^{-\pi\,i/3}\,
\underline{\alpha}^{(2/3)}.\,\underline{D}^{2/3}
+\beta^{(2/3)}\,m^{2/3}\right)^{\!3}\!\psi$
\vskip .1truein
$= \ds\Bigg.\; \left( \ba{l}\ds-\,\sum_{i,j,k}
\alpha^{(2/3)}_i\alpha^{(2/3)}_j\alpha^{(2/3)}_k 
  D^{2/3}_i D^{2/3}_j D^{2/3}_k \\\;+\;\; 
  \ds e^{-2\,\pi\,i/3}\,m^{2/3}\,\sum_{i\,,\,j} \left(\ba{l}
  \quad\alpha^{(2/3)}_i\alpha^{(2/3)}_j\beta\\
  +\;\alpha^{(2/3)}_i\beta\quad\;\;\;\alpha^{(2/3)}_j\\
  +\;\beta\quad\;\;\;\alpha^{(2/3)}_i\alpha^{(2/3)}_j \ea\right)
  \!D^{2/3}_i D^{2/3}_j
  \\\;+\;\; 
  \ds e^{-\pi\,i/3}\,m^{4/3} \left( \ba{l}
  \quad\underline{\alpha}^{(2/3)} \left[\beta^{(2/3)}\right]^{\!2}\\
  +\;\beta^{(2/3)}\,\underline{\alpha}^{(2/3)}\beta^{(2/3)}\\
  +\left[\beta^{(2/3)}\right]^{\!2}\underline{\alpha}^{(2/3)} \ea\right)
  \!\!.\,\underline{D}^{2/3}\\\;+\; 
  \left[\beta^{(2/3)}\right]^3 \!\!m^2 
\ea \right)\!\psi$
\vskip .1truein
$= \ds\Bigg.\; \left( \ba{l}\ds-\;\frac{1}{3!}\sum_{i,j,k}
  \left\{\alpha^{(2/3)}_i D^{2/3}_i\,,\, \alpha^{(2/3)}_j D^{2/3}_j\,,\,
  \alpha^{(2/3)}_k D^{2/3}_k \right\}\\\ds\Bigg.+\;\; 
  \ds e^{-2\,\pi\,i/3}\,m^{2/3}\,\frac{1}{3!}\sum_{i\,,\,j}
  \left\{\alpha^{(2/3)}_i D^{2/3}_i,\,\alpha^{(2/3)}_j
    D^{2/3}_j,\,\beta\right\}\\\ds\Bigg.+\;\;
  e^{-\pi\,i/3}\,m^{4/3}
  \left\{\underline{\alpha}^{(2/3)},\,\beta\,,\,\beta\right\}
  .\,\underline{D}^{(2/3)}\\\ds\Bigg.+\; 
  \left[\beta^{(2/3)}\right]^3 \!\!m^2 
\ea \right)\!\psi$
\vskip .1truein
$= \ds\Bigg.\; \left( -\,\left[\underline{\alpha}^{(2/3)}\right]^3
  \!\!\!.\,\underline{D}^2 \,+\, \left[\beta^{(2/3)}\right]^3
  \!\!m^2 \right)\!\psi \;\;\equiv\; 
\left( -\,\nabla^2 + m^2\right)\!\psi \\\\
\Rightarrow  \left\{\ba{l}
\ds\Bigg.
\left\{\alpha^{(2/3)}_i,\,\alpha^{(2/3)}_j,\,\alpha^{(2/3)}_k\right\} = 
3!\;\delta_{ijk}\,\id \quad,\quad\\
\ds\Bigg.
\left\{\alpha^{(2/3)}_i,\,\alpha^{(2/3)}_j,\,\beta^{(2/3)}\right\} = \zero
\quad,\quad
\left\{\alpha^{(2/3)}_i,\,\beta^{(2/3)}\,,\,\beta^{(2/3)}\right\} = \zero
\quad,\quad\\
\ds\Bigg.\left[\alpha^{(2/3)}_i\right]^3 = \id
\quad,\quad \left[\beta^{(2/3)}\right]^3 = \id
\ea\right.$
\vskip .1truein

For general $p$, Dirac Algebra is analytic continued to
\be\ba{l}\ds\Bigg.\left\{\alpha^{(2/p)}_{i_1},\,\alpha^{(2/p)}_{i_2},\dots,\,
\alpha^{(2/p)}_{i_{p-2}},\,\alpha^{(2/p)}_{i_{p-1}},\,
\alpha^{(2/p)}_{i_p}\right\} = p!\;\,\delta_{\;i_1 i_2 \,\cdots \,i_p} 
\,\id\quad,\quad\\
\ds\Bigg.\left\{\alpha^{(2/p)}_{i_1},\,\alpha^{(2/p)}_{i_2},\dots,\,
\alpha^{(2/p)}_{i_{p-2}},\,\alpha^{(2/p)}_{i_{p-1}},\,
\beta^{(2/p)}\right\} = \zero\quad,\quad\\
\ds\Bigg.\left\{\alpha^{(2/p)}_{i_1},\,\alpha^{(2/p)}_{i_2},\dots,\,
\alpha^{(2/p)}_{i_{p-2}},\,\beta^{(2/p)}\,,\,\beta^{(2/p)}\right\} = 
\zero\quad,\quad\\
\qquad\qquad\qquad\vdots\\
\ds\left\{\Big.\right.\!\alpha^{(2/p)}_{i_1},\,
\underbrace{\beta^{(2/p)}\,,\dots,\,\beta^{(2/p)}}_{(p-1)\mbox{-times}}
\!\!\left.\Big.\right\} = \zero\quad,\quad\\
\\
\ds \left[\alpha^{(2/p)}_i\right]^p = \id\quad,\quad
\left[\beta^{(2/p)}\right]^p = \id
\ea\ee

From here, we may proceed on to find representations of this analytic
continued algebra and study the properties of the associated analytic
continued spinors. Perhaps they have interesting properties.

As a hint, even the matrix representation of the finite difference of
$D^s$ itself has surprising properties. To this we turn to next.

\section{Application: Matrix Representation $-$\\
{\normalsize\bf Analytic continuation of Matrices,\\
 from local Finite Difference to
 non-local Finite Difference}\label{sec_fin_diff}}

\begin{quote}{\em ``We [he and Halmos] share a philosophy about linear
    algebra: we think basis-free, we write basis-free, but when the
    chips are down we close the office door and compute with matrices
    like fury.''} --- Irving Kaplansky in {\em Paul Halmos: Celebrating
    50 Years of Mathematics}\end{quote}

Given a generic matrix $[M]$, we all know how to compute $[M]^n$, the
matrix $[M]$ raised to an integer power $n\in\z$. It is just trivially
multiplying the matrix $[M]$ by itself $n$-times.

Now we wish to compute $[M]^{\sigma}$, the matrix $[M]$ raised to a
real non-integer power $\sigma\in\r$.  A generic matrix may have
degenerate eigenvalues and so cannot in general be diagonalized.
However, we can obtain $[M]^{\sigma}$ as follows:

For rational $\sigma = p/q$, where $p,\,q\in\z$,
  
$[M]^{p/q}$ can be obtained by solving for each element of the matrix
 $[A]$ in\\
the matrix equation$\Bigg.\quad [A]^q = [M]^p\;$\\
since formally $\Bigg.\quad([M]^{p/q})^q =
\underbrace{[M]^{p/q}\,[M]^{p/q}\cdots[M]^{p/q}}_{q
\mbox{\footnotesize -times}} = [M]^p = [A]^q\quad$\\
and so$\quad[A] = [M]^{p/q}$.

In Finite Difference, if we choose the matrix representations of
differentiation $D^1$ to be
\be [D^1_{\!x}] = \left(\ba{rrrrrr}
1&\sm1\\
 &1&\sm1\\
 & &\!\ddots\!\\
 & & &\;1&\sm1\\
 & & &   &1
\ea\right)\!\Big/(\Delta x)\ee
with
\be [D^2_{\!x}] = [D^1_{\!x}] \, [D^1_{\!x}] = [D^1_{\!x}]^2 =
\left(\ba{rrrcrr}
1&\sm2&1\\
 &1   &\sm2&\;1\\
 &    &\!\ddots\!\\
 &    &    &\;1&\sm2&1\\
 &    &    &   &1   &\sm2\\
 &    &    &   &    &1
\ea\right)\!\Big/(\Delta x)^2\ee
and that of integration $D^{-1}$ as the inverse of $D^1$,
\be [D^{-1}_{\!x}] = [D^1_{\!x}]^{-1} =
\left(\ba{ccccc}
1&1&1&1&\!\cdots\!\\
 &1&1&1&\!\cdots\!\\
 & &\!\ddots\!\\
 & & &1&1\\
 & & & &1
\ea\right)\!(\Delta x)\ee
we then have
$$[D^m_{\!x}] = [D^1_{\!x}]^m = \overbrace{[D^1_{\!x}] \, [D^1_{\!x}]
  \cdots [D^1_{\!x}]}^{m \mbox{\footnotesize-times}}$$
$$[D^{-m}_{\!x}] = [D^1_{\!x}]^{-m} =
\underbrace{[D^1_{\!x}]^{-1} \, [D^1_{\!x}]^{-1} \cdots
  [D^1_{\!x}]^{-1}}_{m \mbox{\footnotesize-times}}$$
$$[D^1_{\!x}]\,[f(x)] \,=\,\!
\left(\ba{rrrrrr}
1&\sm1\\
 &1&\sm1\\
 & &\ddots\\
 & & &\;1&\sm1\\
 & & &   &1
\ea\right)\!\!\!
\left( \ba{c} \!\!f(x_n)\!\!\\ \!\!f(x_{n-1})\!\!\\ \!\!
    \vdots\!\!\\ \!\!f(x_2)\!\!\\ \!\!f(x_1)\!\!
\ea\right)\!
\Big/(\Delta x) \,=\, \!\left( \ba{c} \!\!D^1_{\!x} f(x_n)
    \!\!\\ \!\!D^1_{\!x} f(x_{n-1}) \!\!\\ \!\!\vdots\!\!\\
    \!\!D^1_{\!x} f(x_2)\!\!\\ \!\!f(x_1)/(\Delta x)
\!\!\ea\right)$$
$$[D^{-1}_{\!x}]\,[f(x)] \,=\,\!
\left(\ba{ccccc}
1&1&1&1&\!\cdots\!\\
 &1&1&1&\!\cdots\!\\
 & &\!\ddots\!\\
 & & &1&1\\
 & & & &1
\ea\right) \!\!\!
\left( \ba{c} \!\!f(x_n)\!\!\\ \!\!f(x_{n-1})\!\!\\ \!\!
    \vdots\!\!\\ \!\!f(x_2)\!\!\\ \!\!f(x_1)\!\!
  \ea \right)
(\Delta x) \,=\, \left( \ba{c}\!\! D^{-1}_{\!x} f(x)
    \big|^{x_n}_{x_1} \Big.\!\!\!\!\\ \!\! D^{-1}_{\!x} f(x)
    \big|^{x_{n-1}}_{x_1} \Big.\!\!\!\!\\ \!\!\vdots\!\!\!\!\\ \!\! 
    D^{-1}_{\!x} f(x)\big|^{x_2}_{x_1}\Big.\!\!\!\!\\ \!\!
    f(x_n)\,(\Delta x) \!\!\!\!
  \ea \right)$$
where $x_k = x_1 + (k-1)\,\Delta x$.

$[D^1_{\!x}]\,$ have degenerate eigenvalues and thus cannot be
diagonalized. However, following the above approach, we can compute
the matrix representation of $D^{\sigma}_{\!x},\;\sigma\in\r$.
It can be verified that
\be [D^\sigma_{\!x}] = \left\{\ba{ll}\left(\ba{lllcl}
b(\sigma,1)&b(\sigma,2)&b(\sigma,3)&\!\cdots\!&b(\sigma,n)\\
 &b(\sigma,1)&b(\sigma,2)&\!\cdots\!&b(\sigma,n\sm1)\\
 & &b(\sigma,1)&\!\cdots\!&b(\sigma,n\sm2)\\
 & & &\!\ddots\!\\
 & & & &b(\sigma,1)\ea\right)\!
\Big/(\Delta x)^\sigma &\mbox{for}\quad\sigma>0\\
\\
\left(\ba{lllcl}
b(\sm\sigma,1)&b(\sm\sigma,2)&b(\sm\sigma,3)&\!\cdots\!&b(\sm\sigma,n)\\
 &b(\sm\sigma,1)&b(\sm\sigma,2)&\!\cdots\!&b(\sm\sigma,n\sm1)\\
 & &b(\sm\sigma,1)&\!\cdots\!&b(\sm\sigma,n\sm2)\\
 & & &\!\ddots\!\\
 & & & &b(\sm\sigma,1)\ea\right)^{\!\!-\mbox{1}}
\!\!\!\!\!(\Delta x)^{-\sigma} &\mbox{for}\quad\sigma<0\ea\right.
\label{D_matrix_rep}\ee
satisfy
\be [D^\sigma_{\!x}] = [D^{\sigma_1}_{\!x}]\,[D^{\sigma_2}_{\!x}]
\cdots [D^{\sigma_p}_{\!x}]\qquad\mbox{for}\quad\sigma = 
\sigma_1+\sigma_2+\cdots+\sigma_p\ee
where $\ds\;\;b(\sigma,k) = \frac{(-1)^{(k-1)}}{\Gamma(k)}
\lim_{\epsilon\to 0}\frac{\Gamma(1\!+\!\sigma\!+\!\epsilon)}
{\Gamma(2\!+\!\sigma\!+\!\epsilon\!-\!k)}\;$ 
which is incidentally the $k$-th of the binomial expansion 
$\bigg.(1+(-1))^\sigma$.

The matrix representation $[D^m]$ is {\em sparse} while $[D^\sigma]$
is in general {\em dense} --- all the elements in the upper
tri-diagonal block become non-zero.

In Finite Difference, sparse matrix entails taking the differences
between only neighboring sets of points, while dense matrix entails
taking the differences among points almost everywhere in the domain
--- a {\em non-local effect}.

From (\ref{D_matrix_rep}),
\be D^\sigma f(x_m) \equiv \lim_{n\to\infty}
[D^\sigma]\,[f(x)]\bigg|_{x\,=\,x_m}\!\!\!\!\!=\lim_{n\to\infty}
\sum_{k=1}^m b(\sigma,k)\,f(x_{m-k+1})\;(\Delta x)^{-\sigma}\ee
where $1\ll m\le n$.

If $f(x)$ is an integer power series, $D^\sigma f(x_m)$ on l.h.s. is
in general a non-integer power series. The corresponding matrix
reprensentation on {r.h.s.} is a sum of ordinary integer power series.
The non-local effect can be then seen to arise from approximating the
non-integer power series by a sum of ordinary integer power series.

This is in parallel with the application of fractional derivative as a
pseudo-differential operator in non-local field theory by Barci {\it et
  al} \cite{non_local_field_th}.

For the case of $[M]^u$ and $[M]^s$, the matrix $[M]$ raised to
irrational $u$ and complex $s$ respectively, we turn to the series
expansion method in equation (\ref{op_series}) below.

\section{Analytic continuation of Generic Operators\label{sec_op_series}}

Now, let's go beyond the analytic continuation of differential and
integral operators to the analytic continuation of generic operators.

We are used to thinking of an operator acting once, twice, three
times, and so on. However, an operator acting integer times can
be analytic continued to an operator acting complex times by making
the following observation:

A generic operator $A$ acting complex $s$-times can be formally
expanded into a series as
\bea 
A^s &=& \left( w\Big.\id - \Big[ w\id - A \Big]
   \right)^{\!s} \;=\;w^s\!\left(\id-\Big[\id-\frac{1}{w}\,A\Big]
\right)^{\!\!s}\nn\\ 
&=& w^s \left(\id + \sum_{n=1}^\infty \frac{(-1)^n}{n!} \left[
    \prod_{k=0}^{n-1} (s \!-\! k) \right] \!\left[ \Big.\id -
    \frac{1}{w}\,A \right]^{\!n} \right) \nn\\ 
&=& w^s\left( \id + \sum_{n=1}^\infty \frac{(-1)^n}{n!} \left[
    \prod_{k=0}^{n-1} (s \!-\! k) \right] \!\!\left[\,\id + \sum_{m=1}^n 
     \!\left(\frac{-1}{w}\right)^{\!\!m} \!\! {n \choose m} A^m \right]
      \right)
\label{op_series}
\eea
where $\;s,w\in\c$, and $\;\id\;$ is the identity operator.

In the nested series on {r.h.s}, all the operators $A$'s are raised
to integer powers which we can evaluate as usual.  The region of
convergence in $\,s\,$ and the rate of convergence of the series will
in general be dependent on operator $A,\;$ parameter $w,\;$ and the
operand on which $A$ acts.  The resulting series then defines $A^s$,
the analytic continuation of the operator $A$, in the region of $s$
where it converges.

\section{Problems and Challenges\label{sec_problems}}
\begin{quote}``Mathematics is not yet ready for such problems,'' --- Paul
  Erd\"{o}s in {\it The American Mathematical Monthly}, Nov.
  (1992)\end{quote}

\subsection{Analytic continuation of Bernoulli numbers and
  polynomials, a new formula for the Riemann zeta function, and the
  Phenonmenon of Scattering of zeros}

Examples of interesting mathematical applications are analytic
continuation of Bernoulli numbers and polynomials, the derivation of a
new formula for the Riemann zeta function in terms of a nested series
of Bernoulli numbers, and the observation of particle-physics-like
scattering phenomenon in the zeros of the analytic continued
polynomials as described in \cite{woon_bernoulli_con}.

For instance, an operator was found in \cite{woon_tree} to
generate Bernoulli numbers.  Applying the series expansion to the
operator analytic continues the Bernoulli number to a function
\be B(s) = w^s\;\Gamma(1+s)\left( \frac{1}{2} +\sum_{n=1}^\infty
  \frac{(-1)^n}{n!}\!\left[ \prod_{k=1}^n (s \!-\! k) \right]\!\!
  \left[\frac{1}{2}+\!\sum_{m=1}^n \left(\frac{-1}{w}\right)^{\!\!m}\!\!
    {n\choose m}\frac{B_{m+1}}{(m\!+\!1)!}\right] \right)
\label{B_series}\ee
which was verified to converge, for $\;$\re$(s) > (1/w)\,,\;$ real $\;w
 > 0\,,\,$ to
\be B(s) = -\;s\;\,\zeta(1-s) \label{B_zeta}\ee
\begin{figure}[hbt]
\begin{center}
\begin{tabular}{c}
\psfig{figure=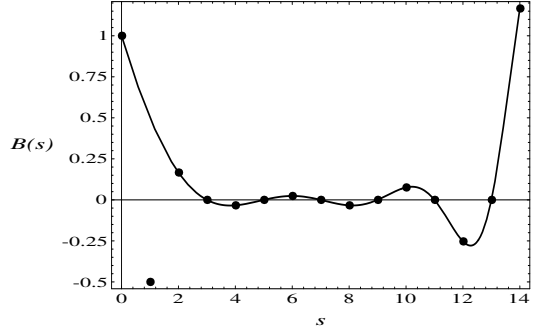,height=125pt,width=200pt}
\end{tabular}
\caption{The curve $\,B(s)\,$ runs through the points of all $\,B_{n}$
  except $\,B_1$.\label{B(s)_curve}}
\end{center}
\end{figure}

Note that $\quad B(n)\,=\,B_n\;$ for $\;n\ge 2\quad$
but $\quad B(1) = 1/2\;$ while $\;B_1 = -\,1/2$.\\
It was then realised that there is actually an arbitrariness in the
sign convention of $B_1$. The analytic continuation of Bernoulli
numbers fixes the arbitrary sign convention, and requires
the generating function of Bernoulli numbers to be redefined for
consistency as
\be\frac{z}{e^z-1}\,=\,\sum_{n=0}^\infty (-1)^n
\frac{B_n}{n!}\,z^n\;,\quad |z|<2\pi\;,\;\;n\in\mbox{\z}^+\label{B_newdef}\ee
or
\be B_n\,=\,\frac{(-1)^{n+1}}{n+1}\,\sum_{k=0}^{n-1}\,(-1)^k\, 
{n\!+\!1\choose k}\,B_k\;,\quad B_0 = 1\label{B_newdef_recursion}\ee
which only changes the sign in the conventional definition
\cite{math_defn} of the only non-zero odd Bernoulli
number, $B_1$, from $\;B_1 = -\,1/2\;$ to $\;B_1 = B(1) = 1/2\;$.

From (\ref{B_series}) and (\ref{B_zeta}), by the functional equation
of the Riemann zeta function (\ref{zeta_func_eqn}),
$$ \!\!\!\!\!\!\!\!\!\!\zeta(s) = -\; \frac{(2\pi w)^s}{2} 
  \lim_{\;{\ds \hat{s}}\to {\ds s}} \left\{\!
  \frac{ \left( \ds \frac{1}{2} + \sum_{n=1}^\infty \frac{(-1)^n}{n!}
  \left[ \prod_{k=1}^n (\hat{s} \!-\! k) \right] \left[ \frac{1}{2} +
  \sum_{m=1}^n \left(\frac{-1}{w}\right)^{\!\!m} \!\!  {n \choose m}
  \frac{B_{m\!+\!1}}{(m+1)!} \right] \!\right)}
{\ds \cos\left(\frac{\pi \hat{s}}{2}\right)} \right\} $$
\be = -\; \frac{(2\pi w)^s}{2} 
  \lim_{\;{\ds \hat{s}}\to {\ds s}} \left\{\!
  \frac{ \left( \ds \frac{1}{2} + \sum_{n=1}^\infty \frac{(-1)^n}{n!}
  \left[ \prod_{k=1}^n (\hat{s} \!-\! k) \right] \left[ \frac{1}{2} -
  \sum_{m=1}^n \left(\frac{-1}{w}\right)^{\!\!m} \!\!  {n \choose m}
  \frac{\zeta(-m)}{m!} \right] \!\right)}
{\ds \cos\left(\frac{\pi \hat{s}}{2}\right)} \right\}
  \label{zeta(s)_zeta(-n)} \ee
a nested sum of the Riemann zeta function itself evaluated at
negative integers, which converges for \re$(s) > (1/w)\;,\;$ real 
$\; w > 0 \;,\;$ with redefined $B_1 = 1/2$, and the limit only
needs to be taken when $\;s = 1,3,5,\dots \in$ \z${}^+_{odd}$, the set
of positive odd integers, for which the denominator
$\;\ds\cos\left(\frac{\pi s}{2}\right) = 0$.

\subsection{Quantum Operators}
\begin{quote}{\em ``The modern physicist is a quantum theorist on
    Monday, Wednesday, and Friday and a student of gravitational
    relativity theory on Tuesday, Thursday, and Saturday. On Sunday he
    is neither, but is praying to $\dots$ find the reconciliation
    between the two views.''}\\ --- Norbert Wiener\end{quote}

Quantum Mechanics, Quantum Field Theories and Canonical Quantum
Gravity are full of non-commutative operators.

As a start, let's try to apply the idea of analytic continuation of
operators to the creation and annihilation operators of a simple
harmonic quantum oscillator \cite{qm}
\be a^\dag\,|n\big>\;=\,\sqrt{n\!+\!1}\;\,|n\!+\!1\big>\,,\quad
a\,|n\big>\;=\,\sqrt{n}\;\,|n\!-\!1\big>\,,\quad a\,|0\big>\;=\,0
\label{def_a}\ee
From the nested series expansion (\ref{op_series}),
\bea a^s_{(w)}&=&\left(\Big.w\id-\left[\big.w\id-a\right]
\right)^{\!s} \nn\\
&=&w^s\left(\id+\sum_{p=1}^\infty \frac{(-1)^p}{p!}\left[
    \prod_{k=0}^{p-1} (s\!-\!k)\right]\!\!\left[\,\id+\sum_{m=1}^p
    \left(\frac{-1}{w}\right)^{\!\!m}\!{p\choose m}\,a^m\right]
\right)\nn\eea 
and similarly for $\;a^{\dag\,s}_{(w)}$.

Consider $a^s$ acting on the quantum state $|n\big>$,
\be\!\!\!\!\!\!\!\! a^s_{(w)}|n\big>\,=
w^s\left(\,|n\big>+\!\sum_{p=1}^\infty \frac{(-1)^p}{p!}\left[
\prod_{k=0}^{p-1} (s\!-\!k)\right]\!\!\left[\,|n\big>+
\!\sum_{m=1}^{p\le n}\left(\frac{-1}{w}\right)^{\!\!m}\!{p\choose m}\,
\!\sqrt{\!\frac{n!}{(n\!-\!m)!}}\;|n\!-\!m\big>\right]
\right)\label{a_s_on_state}\ee
\be\Rightarrow\quad a^s_{(w)}\big|n\big>\,=\,\sum_{m=0}^{n}
w_{n\!-\!m}(s,n,w)\,\big|n\!-\!m\big>\ee
  
The nested series expression is straightforwardly computable but the
interpretation is not clear. 

When $s$ is not an integer, the inner series nested within the outer
series on {r.h.s.} of (\ref{a_s_on_state}) terminates at the
${(p\!=\!n)}$-th term.  However.  the outer series does not terminate
but is an infinite sum for both operators. On the other hand, the
eigenvalues of the creation operator for simple harmonic quantum
oscillator are not bounded above.  under the action of
$a^{\dag\,s}_{(w)}$, $|n\big>$ seems to transform into a superposition
of $|n\big>$ and an infinite towers of all other states higher than
$|n\big>$.

\noindent{\em Case\ } $s\!=\!k\in\mbox{\sf Z}^+,\;w\in\mbox{\sf R}
\mbox{ or }\mbox{\sf C}\,:$
\begin{equation} w_{m}(k,n,w)\,= \left\{\begin{array}{ll}
\!\!\!\sqrt{n!/(n-k)!}&(m\!=\!n\!-\!k)\\
0&\mbox{otherwise}\end{array}\right.\,,\end{equation}
and so we recover (\ref{def_a}) from the nested series expansion
(\ref{a_s_on_state}) independent of $w$.\\\par

\noindent{\em Case\ } $s\in\mbox{\sf R},\;s\not\in\mbox{\sf Z}^+,\;
w\in\mbox{\sf R}\mbox{ or }\mbox{\sf C}\,:$\\
There may be divergences in the nested series expansions. Much
remains to be worked out and clarified.

Now, we turn to the commutators of these analytic continued operators. 

A formal nested series expansion of the commutators of $a^s_{(w)}$ and
$a^{\dag\,s'}_{(w)}$ is

$[\;a^s_{(w)},\,a^{\dag\,s}_{(w')}]$
\vskip .1truein
$=\left[\ba{l}\ds w^s\left(\id+\sum_{p=1}^\infty \frac{(-1)^p}{p!}\left[
\prod_{k=0}^{p-1} (s\!-\!k)\right]\!\!\left[\,\id+\sum_{m=1}^p
\left(\frac{-1}{w}\right)^{\!\!m}\!{p\choose m}\,a^m\right]\right)\,,\\
\ds w'^s\left(\id+\sum_{p'=1}^\infty \frac{(-1)^{p'}}{p'!}\left[
\prod_{k'=0}^{p'-1} (s'\!-\!k')\right]\!\!\left[\,\id+\sum_{m\!'=1}^{p'}
\left(\frac{-1}{w'}\right)^{\!\!m\!'}\!{p'\choose m'}\,
a^{\dag\,m\!'}\right]\right)\ea\right]$
\vskip .1truein
\be=w^s {(w')}^{s'}\left(\ba{l}\ds\sum_{p,\,p'=1}^\infty\!
\frac{(-1)^{p+p'}}{p!\,p'!}\!
\left[\prod_{k=0}^{p-1} (s\!-\!k)\right]\!\!
\left[\prod_{k'=0}^{p'-1} (s'\!-\!k')\right]\\
\ds\quad\left(\sum_{m=1}^p\sum_{m\!'=1}^{p'}
\frac{(-1)}{w^m w^{m\!'}}{\bigg.}^{\!\!\!\!\!m\!+\!m\!'}\!\!
{p\choose m}\!{p'\choose m'}\,[\,a^m,\,a^{\dag\,m\!'}]\right)\ea\right)
\label{aadag_commutator}\qquad\qquad\qquad\ee

From the canonical commutation relations \cite{qft},
\be [\;a\,,\,a^\dag\,]\,=\,1\,,\qquad
[\;a\,,\,a\,]\,=\,0\,=\,[\;a^\dag,\,a^\dag\,] \ee
the $[\,a^m,\,a^{\dag\,\!m\!'}]$ in the nested series can be
evaluated as usual, and thus the nested series expansion
(\ref{aadag_commutator}) is formally computable.

Similar generalization applies to fermionic operators satisfying
Grassmann algebra, SUSY operators
\begin{eqnarray}\{Q\,,Q^\dag\}&=&2H/\omega\,\leadsto\,
\big\{Q^s_{(w)},Q^{\dag\,s\!'}_{(w')}\big\}\,,\end{eqnarray}
Virasoro generators \cite{virasoro} in String theories
\begin{eqnarray}[L_n,L_m]&=&(n\!-\!m)L_{n\!+\!m}+
\frac{c}{12}n(n^2\!-\!1)\,\delta_{n,\!-\!m}\\
&\leadsto&\big[L^s_{n(w)},L^{s\!'}_{m(w')}\big]\,,\nonumber\end{eqnarray}
Superconformal algebra \cite{superstring} in Superstring, their
respective vertex operators, and Lie algebra in general.

These generalizations seem to have interesting mathematical
structures. Further aspects and detailed computations will be
presented and explored in a forthcoming paper \cite{woon_SUSY_con}.

\section{Conclusion}

Analytic continued operators have been demonstrated to exhibit
intriguing properties. In addition, fractional derivatives in the
conventional Riemann-Liouville Fractional Calculus do not generally
commute but an extension in which they commute has been found and
applied to various fields. These methods of analytic continuation of
operators may after all turn out to be a general and powerful
exploration tools in Maths, Physics, Sciences, and Engineering.

Calculus is never quite the same again. It would be interesting to
imagine what Newton and Leibniz would say on this analytic
continuation of their discoveries --- Calculus, and Dirac of his
equation -- Dirac equation.

Perhaps the most unexpected, and yet ``inconsequential'', consequence
is that Figure \ref{B(s)_curve} clearly points out that the commonly
adopted definition of the 1st Bernoulli number $\,B_1\,$ has the {\em
  wrong}$\,$ sign. There was actually arbitrariness in its sign convention
and the analytic continuation of the operator that generates Bernoulli
numbers \cite{woon_bernoulli_con} fixes that arbitrariness, requiring
that $\,{B_1 = -1/2}\,$ to be redefined as $\,{B_1 = 1/2}\,$ for consistency.
However, the $B_1 = 1/2$ definition has been so widely used --- in
every Math, Physics and Engineering book or paper where Bernoulli
numbers appear, one almost certainly find $B_1 = -1/2$. I can only
hope that the readers will be persuaded in the light of this new
mathematical fact to change and adopt the consistent definition $B_1 =
1/2$ and the corresponding defining equations (\ref{B_newdef}) and
(\ref{B_newdef_recursion}).

\vskip .2truein

\noindent {\large \em Acknowledgement}

Special thanks to V. Adamchik, D. Bailey, W. Ballman, J. Borwein, P.
Borwein, P.  D'Eath, U. Dudley, C. Isham, K. Odagiri, Y.L. Loh, B.
Lui, H. Montgomery, A. Odlyzko, S. Shukla, I.N. Stewart, M. Trott, and
B. Wandelt for discussion, all the friends in Cambridge for
encouragement, and Trinity College UK Committee of Vice-Chancellors
and Principals for financial support.

\end{document}